\documentclass[10pt,a4paper]{article}

\usepackage[T1]{fontenc}
\usepackage[utf8]{inputenc}
\usepackage{lmodern}
\usepackage{microtype}
\usepackage[margin=2.4cm]{geometry}
\usepackage{amsmath,amssymb,amsthm}
\usepackage{graphicx}
\usepackage{booktabs}
\usepackage{setspace}
\usepackage[labelfont=bf,font=footnotesize]{caption}
\usepackage[numbers,sort&compress]{natbib}
\usepackage{xcolor}
\usepackage{xurl}
\usepackage[colorlinks=true,linkcolor=black,citecolor=blue!50!black,
            urlcolor=blue!50!black]{hyperref}

\usepackage{mathtools}
\usepackage{bm}
\usepackage{units}
\graphicspath{{figures/}}

\newcommand{\PaperTitle}{A hybrid wave-mode formulation for the vibro-acoustic analysis of 2D periodic structures}
\newcommand{\PaperAuthors}{C. Droz$^{1,2}$, C. Zhou$^{1}$, M.N. Ichchou$^{1}$ and J.-P. Lain\'e$^{1}$}
\newcommand{\PaperAffil}{%
  $^{1}$Ecole Centrale de Lyon, 36 Avenue Guy de Collongue, 69134 Ecully Cedex, France\\
  $^{2}$Airbus Helicopters, A\'eroport Inter. Marseille-Provence, 13725 Marignane Cedex, France}
\newcommand{\PaperDate}{Accepted manuscript, originally 2016}
\newcommand{\PaperKeywords}{Wave, finite element, periodic structure, reduced model, stiffened plate}

\newcommand{\PaperDOI}{10.1016/j.jsv.2015.11.003}
\newcommand{\HalID}{hal-03372965}
\newcommand{\ArxivID}{}          % ex. 2609.01234 --- laisser vide si inconnu

\newcommand{\CiteAs}{%
  C. Droz, C. Zhou, M.N. Ichchou, J.-P. Lainé, \emph{\PaperTitle}, Journal of Sound and Vibration 363 (2016) 285-302, doi:10.1016/j.jsv.2015.11.003}
\makeatletter
\newcommand{\HasDOI}{%
  \ifx\PaperDOI\@empty\expandafter\@secondoftwo\else\expandafter\@firstoftwo\fi}
\newcommand{\HasArxiv}{%
  \ifx\ArxivID\@empty\expandafter\@secondoftwo\else\expandafter\@firstoftwo\fi}
\makeatother

\providecommand{\doi}[1]{\href{https://doi.org/#1}{doi:#1}} 

\begin{document}

% Title page: bold title; authors; copyable citation.
% Affiliations, date, HAL/DOI/AAM live in footnotes (fnsymbol, then arabic resumes).
\begingroup
\renewcommand{\thefootnote}{\fnsymbol{footnote}}
\begin{center}
{\LARGE\bfseries \PaperTitle\par}
\vspace{0.7em}
{\normalsize \PaperAuthors\footnotemark[1]\footnotemark[2]\par}
\end{center}
\footnotetext[1]{\mbox{}\\[-0.7em]\raggedright\PaperAffil}
\footnotetext[2]{\PaperDate.
  Author's version with AI-assisted typesetting and layout. Not the publisher version of record.
  \HasDOI{Definitive publication: \doi{\PaperDOI}.}{No journal DOI yet.}
  HAL: \href{https://hal.science/\HalID}{\HalID}.
  \HasArxiv{arXiv: \href{https://arxiv.org/abs/\ArxivID}{\ArxivID}.}{}}
\endgroup
\setcounter{footnote}{0}

\vspace{0.85em}
{\small
\noindent\textit{How to cite this version.}\\[0.35em]
\raggedright \CiteAs
\par}
\vspace{1.05em}

\begin{abstract}
\noindent
In the framework of vibrational analysis of 2D periodic waveguides, Floquet--Bloch theorem is widely applied for the determination of wave dispersion characteristics. In this context, the Wave Finite Element Method (WFEM) combines Periodic Structure Theory (PST) with standard FE packages, enabling wave dispersion analysis of waveguides involving structurally realistic unit-cells. For such applications, the computational efficiency of the WFEM depends on the choice of the formulation and can lead to numerical issues, worsened by extensive computational cost. This paper presents a coupled wave-mode approach for the determination of wave dispersion characteristics in structurally advanced periodic structures. It combines two scales of model order reduction. At the unit-cell's scale, Component Mode Synthesis (CMS) provides the displacement field associated with local resonances of the periodic structure, while the free wave propagation is considered using a spectral problem projection on a reduced set of shape functions associated with propagating waves, thus providing considerable reduction of the computational cost. An application is provided for a bi-directionally stiffened panel and the influence of reduction parameters is discussed, as well as the robustness of the numerical results.
\end{abstract}


% ============================================================
% FILL — corps : corps du papier
% ============================================================

\section{Literature survey}
Two-dimensional periodic waveguides such as honeycomb sandwiches, stiffened panels or beam lattices are extensively encountered in the field of automotive and aerospace industry. These structures provide high structural integrity and low weight. Yet, an optimal use in engineering requires their vibro-acoustic characteristics to be known since they are prone to fatigue and fracture issues. They are also likely to generate important acoustic radiation in the medium-frequency range. For this reason, numerical simulations are increasingly needed for the design and optimization processes of automotive and aerospace structures using these materials. In the context of vibro-acoustic analysis of composite waveguides, several methods have been proposed to perform analysis of wave propagation in the low-, mid- or high-frequency range.

In the low frequency range, the vibrational behaviour of periodic structures is governed by the macroscopic, or homogeneous material parameters. The mechanical energy propagates to the entire structure and does not produce wave localization in the periodic cell. Several analytical models can be found in the literature for two dimensional periodic waveguides. Mead \cite{MeadJSV1988} proposed a model for free-wave propagation in a flat plate stiffened using an orthogonal array of beams. The same author \cite{MeadJSV1973} developed later a formulation based on Bloch theorem \cite{Bloch1929} and introduced homogenization techniques and model order reduction for wave propagation in periodic waveguides. In \cite{MejdiAtallaJSV2010, MejdiAtallaIJMS2012}, the authors developed further an analytical formulation based on modal expansion technique to predict the vibro-acoustic of orthogonally stiffened panels. Analytical formulations are accurate at low frequency and can be extended with higher orders of the theory including kinematic assumptions. However it requires high-order derivatives leading to an increased mathematical effort to obtain analytical solutions \cite{SutyrinAM1997}. In \cite{ChronopoulosCS2013}, the authors studied the dynamic response of a sandwich panel using a dynamic stiffness approach and several Equivalent Single Layer (ESL) theories using an homogenized orthotropic honeycomb core. The use of modal reduction can overcome some FEM methods limitations between low- and mid- frequency range, but reaches its limits for composite materials which exhibit high modal densities.

In the high frequency range, the mechanical energy mainly remains in the periodic cell and the diffusion between the structure subsystems is weak. Then, the vibro-acoustic behaviour is governed by the periodic cells dynamics and can be determined from a single periodic cell \cite{RuzzeneSMS2003} using Bloch theorem. The averaged dynamic response of the structure under such excitation can be estimated using Statistical Energy Analysis (SEA) \cite{Lyon1995}. However, SEA method suffers significant limitations in the medium frequency range, considering it cannot provide the response of the long-wavelength components.

In the medium-frequency range, the vibratory energy is partially localized in the periodic cell and propagates through the plate. In the last decade, numerical methods based on multi-modal propagation received a lot of attention. Among them, the Wave Finite Element (WFE) method \cite{MaceJASA2005, MencikEJM2005, Zhong1995} is an effective tool for providing the wave dispersion characteristics of a wide range of waveguides. The main feature of the WFE method is to combine periodic structure theory with commercial finite element packages. Numerous applications were proposed for one-dimensional homogeneous and periodic waveguides \cite{MencikEJM2005, MaceJASA2005, IchchouIJSS2008, HouillonJSV2005, WakiJSV2009a, ChronopoulosCS2014}. A formulation was introduced for two-dimensional waveguides \cite{MaceManconiJSV2008} and applied to periodic structures in \cite{ColletIJSS2011, OrreniusWM2013, Renno2011}. Still, the above-mentioned methods are based on Floquet - Bloch theorem and face some drawbacks due to high computational costs when refined FEM meshes are required. Some numerical issues were investigated for example in \cite{WakiJSV2009b} and a reduction method to compute the response of one-dimensional waveguides under harmonic excitations was proposed in \cite{MencikCMAME2012, Renno2010}. However, there is still a lack of studies in the literature concerning model order reduction strategies for the WFE method in the case of 2D periodic waveguides.

As a consequence, the study is often restrained to lower frequencies due to coarsely meshed models or structural approximations. In \cite{DrozCS2014}, the authors proposed a reduced strategy for studying propagating waves dispersion characteristics in structurally advanced or composite waveguides. Several formulations using Bloch's theorem can be found in the literature for 2D periodic waveguides \cite{LangleyJSV1993, AkroutPhD2005, MaceManconiJSV2008}. These methods require the resolution of an eigenproblem depending on the spatial variables used in the periodicity conditions, and possible degrees of freedom (DOF) condensation. One can discriminate these formulations between "unknown-frequency" and "unknown-wavenumber" resolutions. These different approaches are discussed in Section~\ref{sec2}. Since these formulations generally involve extensive computation when an important degree of discretization is required, the inner degrees of freedom condensation can benefit Component Mode Synthesis (CMS). In \cite{CotoniJSV2008}, the authors successfully applied the CMS to an unknown-frequency WFE method. However, this formulation requires to choose arbitrary real phase constants and is not suited for damped structures. In order to provide a suitable numerical tool for wave analysis in structurally advanced periodic waveguides, this paper focuses on the prediction of propagating waves dispersion characteristics using reduced computation processes. The aim is to provide a suitable numerical tool for computing wave dispersion in two-dimensional periodic systems using a CMS approach for the inner cell DOF condensation, then reducing the spectral eigenproblem projecting the cell's cross-sectional transfer matrix on a reduced set of shape functions. It allows the propagating waves to be determined accurately by solving smaller eigenproblems, and enables low-cost broadband analysis of finely meshed two-dimensional periodic waveguides.

This paper is organized as follows. In Section~\ref{sec2}, some WFE formulations for 2D periodic waveguides are reviewed. The purpose of this section is to classify the methods, highlight their numerical issues and motivate the chosen "unknown-k" formulation. We also introduce the proposed developments presented in the following section. Section~\ref{sec3} describes the proposed reduction strategy. A CMS approach based on mode selection is first developed for the dynamic condensation required in the proposed WFE formulation. A modal analysis is performed on the fixed boundary unit cell; the main contributing modes of the periodic cell are retained to replace the inner cell's DOFs; a dynamic condensation of the modal participations allows to use classical 2D periodic WFE formulation used in \cite{MaceManconiJSV2008}. Then a reduced formulation based on wave interpolation is provided to solve the spectral problem on a broadband frequency range. A subset of solutions is computed at waves cut-on frequencies; a shape basis is determined using a reduced number of propagating waves; an iterative process based on eigenvectors orthogonality is proposed to improve the reduced basis numerical stability; spectral problem is then projected on the reduced-size basis and easily solved for the whole frequency band. Numerical application is brought in Section~\ref{sec4}. Orthogonally stiffened panels are considered; classical WFE solutions are compared with the proposed reduced formulation; the solutions sensitivity to both modal and wave selection is investigated and the computational gains are discussed.

\section{Review of the WFE formulations for 2D periodic waveguides}\label{sec2}
	\subsection{Application of the periodicity relations\label{2.1}}
	In this section, an overview of the WFE formulation is given for the general 2D periodic waveguide. These structures consist of a set of identical and regularly connected elements.
	\begin{figure}[!ht]
  \centering
  \includegraphics[width=0.45\linewidth]{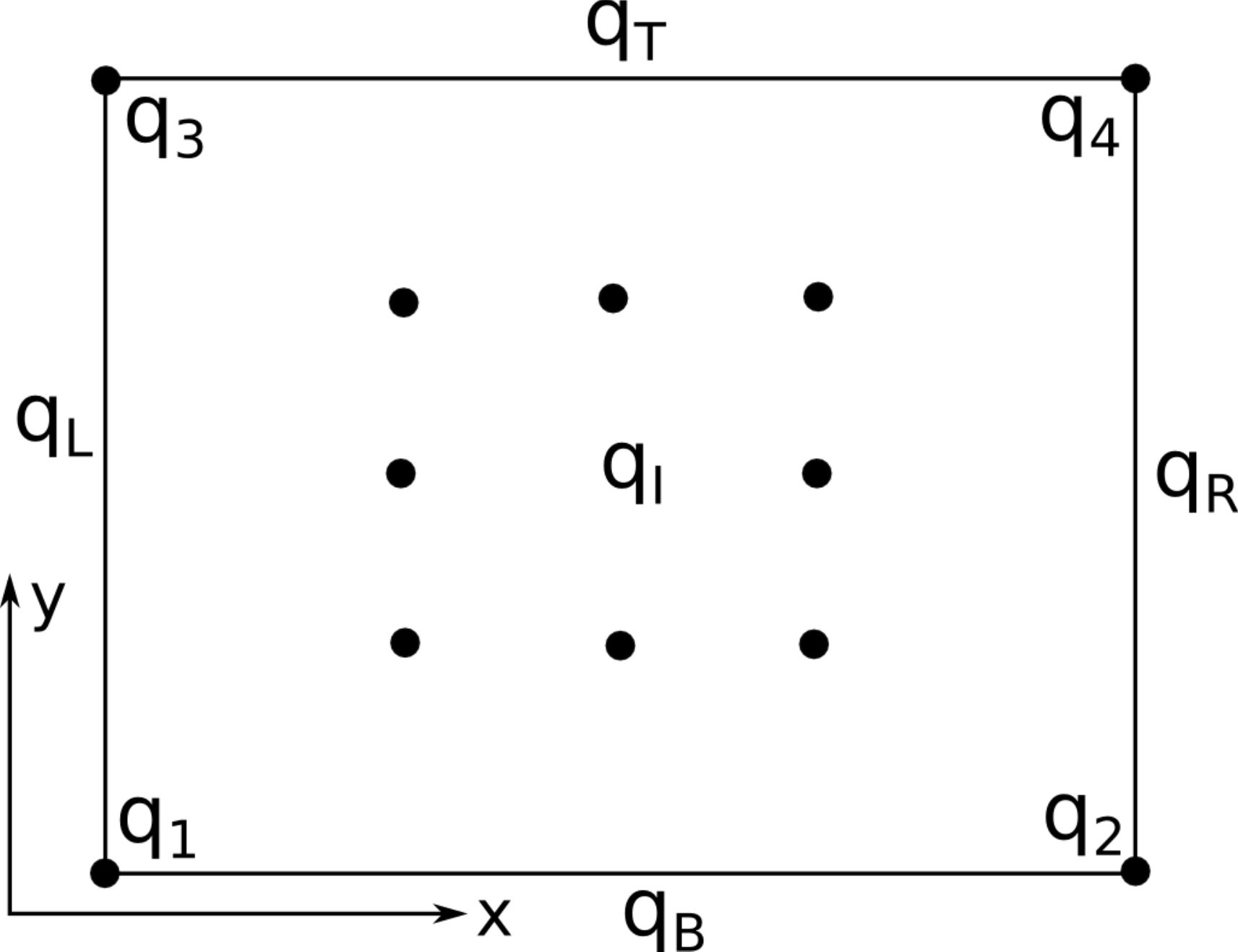}
  \caption{Illustration of the 2D waveguide periodic cell.}
	   \label{periodic_cell}
	\end{figure}
	A periodic cell is shown in Fig.~\ref{periodic_cell}, the degrees of freedom $\mathbf{q}$ are partitioned into edges $(\mathbf{q}_B, \mathbf{q}_T, \mathbf{q}_L, \mathbf{q}_R)$, corners $(\mathbf{q}_1, \mathbf{q}_2, \mathbf{q}_3, \mathbf{q}_4)$ and inner $\mathbf{q}_I$ DOFs. The vector of nodal forces is defined similarly except for the internal DOFs where no external forces are applied: $\mathbf{f}_I =\mathbf{0}$. It is assumed that opposite edges and the four corners have the same number of DOFs. Apart from these conditions, the FEM discretization is arbitrary inside the periodic element.
	The relations between the element displacements can be expressed using Bloch's theorem:
	\begin{equation}
	\mathbf{q}_R=\lambda_x\mathbf{q}_L ,\quad	\mathbf{q}_T=\lambda_y\mathbf{q}_B.
	\end{equation}
	for the edges, and
	\begin{equation}
	\mathbf{q}_2=\lambda_x \mathbf{q}_1 ,\quad \mathbf{q}_3=\lambda_y \mathbf{q}_1 ,\quad \mathbf{q}_4=\lambda_x \lambda_y \mathbf{q}_1,
	\end{equation}
	for the corners. The forces relations can be expressed similarly:
	\begin{equation}
	 \mathbf{f}_R=-\lambda_x \mathbf{f}_L ,\quad \mathbf{f}_T=-\lambda_y \mathbf{f}_B,
	\end{equation}
	for the edges forces, and 
	\begin{equation}
	 \mathbf{f}_2=-\lambda_x \mathbf{f}_1 ,\quad \mathbf{f}_3=-\lambda_y \mathbf{f}_1 ,\quad \mathbf{f}_4=\lambda_x \lambda_y \mathbf{f}_1.
	\end{equation}
	for the corners forces. The propagation constants $\lambda_x = e^{-jk_xd_x}$ and $\lambda_y = e^{-jk_yd_y}$ in $x$ and $y$ directions, involve the wavenumbers $k_x$ and $k_y$ and the element size $(d_x \times d_y)$. The values $k_x d_x$ and $k_y d_y$ are also called \emph{phase constants}. Similarly, the forces equilibrium leads to the following conditions
	\begin{equation}
	\mathbf{f}_B+\lambda_y^{-1} \mathbf{f}_T = \mathbf{0} ,\quad \mathbf{f}_L+\lambda_x^{-1} \mathbf{f}_R = \mathbf{0}
	\end{equation}
	for the edges, and 
	\begin{equation}
	\mathbf{f}_1 \lambda_x \lambda_y + \mathbf{f}_2 \lambda_y + \mathbf{f}_3 \lambda_x + \mathbf{f}_4 = \mathbf{0}
	\end{equation}
	for the corners. 
	
	The cell's equation of motion can be used to form the spectral problem involving the wavenumbers $k_x$, $k_y$ and the frequency. Denoting $\mathbf{K}$ and $\mathbf{M}$ the stiffness and mass matrices and $\omega$ the circular frequency, the governing equation in the element is written, assuming time-harmonic excitation:
	\begin{equation}
	(\mathbf{K} - \omega^2 \mathbf{M})\mathbf{q} = \mathbf{f}
	\label{eqgoverning}
	\end{equation}
		
	Different methods were provided to form a well-conditioned eigenproblem for the determination of wave dispersion characteristics in 2D periodic waveguides. The choice of a method strongly influences the problem size, the results usability and model order reduction possibilities. One can discriminate between a given $(k_x, k_y)$ formulation (see Section~\ref{2.2}) and a given $(\omega, k_x)$ formulation described in Section~\ref{2.3}.

	\subsection{Given $(k_x, k_y)$ formulation\label{2.2}}
	The cell's DOFs can be ordered \cite{LangleyJSV1993} as follows:
	\begin{equation}
	\mathbf{q} = \left[\mathbf{q}_I^T \quad \mathbf{q}_B^T \quad \mathbf{q}_T^T \quad \mathbf{q}_L^T \quad \mathbf{q}_R^T \quad \mathbf{q}_1^T \quad \mathbf{q}_2^T \quad \mathbf{q}_3^T \quad \mathbf{q}_4^T \right]^T
	\end{equation}
	Using the periodicity relations defined Section~\ref{2.1}, one can define the reduced state vector $\mathbf{q}'$ so that $\mathbf{q} = \mathbf{R} \mathbf{q}'$, or:
	\begin{equation}
	\left( \begin{array}{c}
	\mathbf{q}_I\\
	\mathbf{q}_B\\
	\mathbf{q}_T\\
	\mathbf{q}_L\\
	\mathbf{q}_R\\
	\mathbf{q}_1\\
	\mathbf{q}_2\\
	\mathbf{q}_3\\
	\mathbf{q}_4\\
	\end{array} \right) =
	\left(\begin{array}{cccc}
	\mathbf{I} & 0 & 0 & 0\\
	0 & \mathbf{I} & 0 & 0\\
	0 & \mathbf{I}\lambda_y & 0 & 0\\
	0 & 0 & \mathbf{I} & 0\\
	0 & 0 & \mathbf{I}\lambda_x & 0\\
	0 & 0 & 0 & \mathbf{I}\\
	0 & 0 & 0 & \mathbf{I}\lambda_x\\
	0 & 0 & 0 & \mathbf{I}\lambda_y\\
	0 & 0 & 0 & \mathbf{I}\lambda_x\lambda_y
	\end{array} \right)
	\left( \begin{array}{c}
	\mathbf{q}_I\\
	\mathbf{q}_B\\
	\mathbf{q}_L\\
	\mathbf{q}_1
	\end{array} \right)
	\label{eq2.2R}
	\end{equation}
	
\begin{figure}[!ht]
  \centering
  \includegraphics[width=0.45\linewidth]{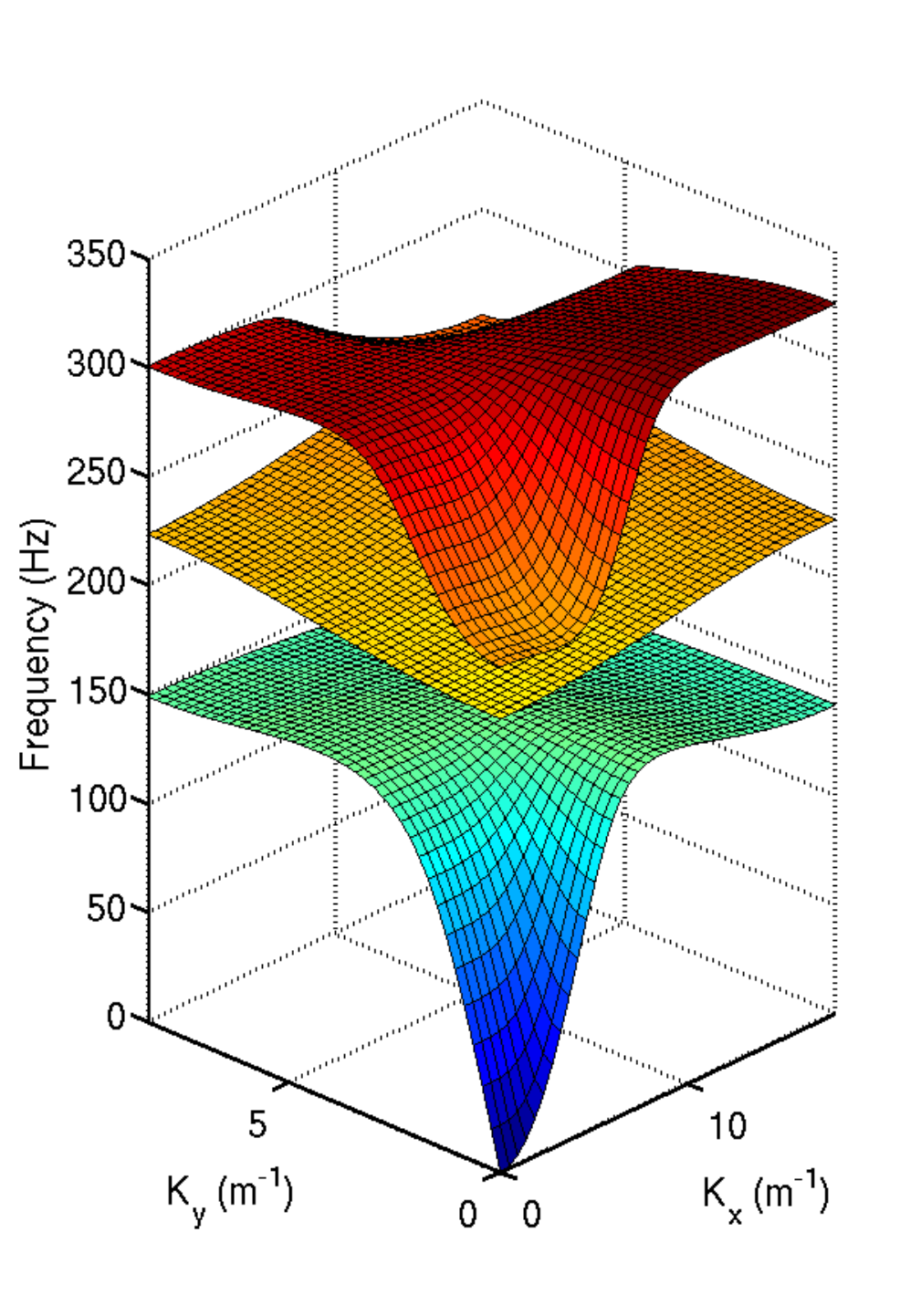}
  \caption{Phase constant surfaces of a bi-directionally stiffened panel, obtained using given-($k_x, k_y$) WFE formulation.}
  \label{Phase_constant_surface}
\end{figure}

	Substituting Eq.~(\ref{eq2.2R}) in the governing equation~(\ref{eqgoverning}) and pre-multiplying by $R^T$ yields:
	\begin{equation}
	\mathbf{R}^T (\mathbf{K} - \omega^2 \mathbf{M}) \mathbf{R} \mathbf{q}' = \mathbf{0}
	\label{eq2.2eig}
	\end{equation}
	In practice, the eigenvalues $\omega_i$ are found for each couple $(\lambda_x,\lambda_y)$ by specifying a range of values for $k_x$ and $k_y$. Then, the propagating solutions can be sorted in order to plot \emph{phase constants surfaces} illustrated in Fig.~\ref{Phase_constant_surface}. One can note that Eq.~(\ref{eq2.2eig}) requires the computation of a sizeable eigenproblem since the inner DOFs $\mathbf{q}_I$ are involved. 
	
	A CMS approach was proposed in \cite{CotoniJSV2008} to reduce the computational expense associated with the phase constant surfaces determination described above. The method uses a subset fixed-interface modes of the periodic element to replace the interior DOFs. The inner DOFs $q_I$ are written as:
	\begin{equation}
	\mathbf{q}_I = \mathbf{\Phi} \mathbf{p}_\varPhi - \mathbf{K}_{\tt II}^{-1} \mathbf{K}_{\tt Ib} \mathbf{q}_b
	\label{eq2.2qI}
	\end{equation}
	where $\mathbf{\Phi}$ is a subset of the clamped boundary local modes, $\mathbf{p}_\varPhi$ contains the associated modal participations, while the term $\mathbf{K}_{\tt II}^{-1} \mathbf{K}_{\tt Ib} \mathbf{q}_b$ stands for the constraint modes where subscript $b$ indicates the periodic element boundary DOFs. The number of modes retained is significantly lower than the number of inner DOFs, the reduced state vector $\mathbf{q}'$ in Eq.~(\ref{eq2.2eig}) being written:
	\begin{equation}
	\tilde{\mathbf{q}}' =	\left( \begin{array}{c}
	\mathbf{p}_\varPhi\\
	\mathbf{q}_B\\
	\mathbf{q}_L\\
	\mathbf{q}_1
	\end{array} \right)
	\end{equation}
	Then, the associated computation time is significantly reduced for each couple of $(\lambda_x,\lambda_y)$ specified. The same authors applied this procedure and obtained significant CPU time reduction in comparison with the standard approach. Notwithstanding, this formulation suffers some drawbacks. Firstly, the specified wavenumbers $(k_x, k_y)$ have to be assumed real hence restraining the study to undamped systems, since no information is given on the imaginary parts which are associated with propagating waves spatial attenuations. Secondly, the solutions provided here are not suited for a further use in forced response analyses, where the frequency is a given parameter. Later formulations of the spectral problem were provided in Section~\ref{2.3}, allowing a dynamic condensation of the element inner DOFs.
		
	\subsection{Given $(\omega, k_x)$ formulation\label{2.3}}
	In \cite{AkroutPhD2005}, the author proceeds a dynamic condensation of the inner DOFs, and gives the explicit coefficients $\mathbf{X}(\omega, \lambda_x), \mathbf{Y}(\omega, \lambda_x), \mathbf{Z}(\omega, \lambda_x)$  so that the spectral problem equation~(\ref{eq2.2eig}) can be written on the quadratic form:
	\begin{equation}
	(\lambda_y \mathbf{X} + \mathbf{Y} + \lambda_y^{-1} \mathbf{Z})	\left( \begin{array}{c}
	\mathbf{q}_L \\
	\mathbf{q}_B \\
	\mathbf{q}_1
	\end{array} \right) = \mathbf{0}
	\label{eq2.3akrout}
	\end{equation}
	
\begin{figure}[!ht]
  \centering
  \includegraphics[width=0.45\linewidth]{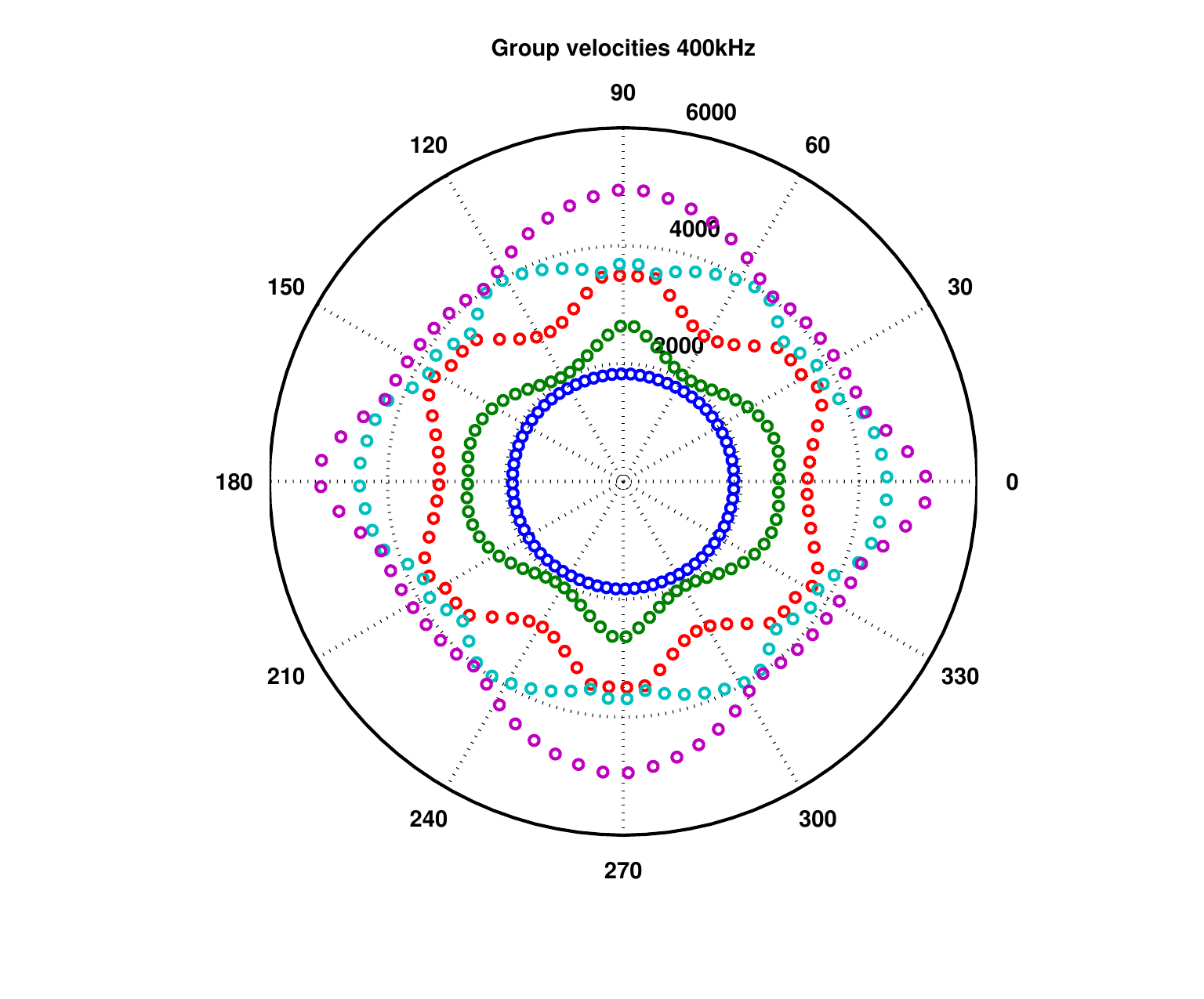}
  \caption{Group velocities of a multi-layered orthotropic plate at 400\,kHz, obtained using given-($\omega, k_y$) WFE formulation and plotted in polar coordinates.}
  \label{fig:wave_directivity}
\end{figure}

	A given $(\omega,k_y)$ formulation can be obtained similarly by inverting the roles of $k_y$ and $k_x$. Further investigations regarding the resolution of linear, quadratic, polynomial or transcendental eigenproblems were carried out in \cite{MaceManconiJSV2008}. The complex solutions $\lambda_y$ are then sorted in order to retain mainly propagating waves. If the wavenumber $k_y$ is purely imaginary the wave is evanescent, if $k_y$ is complex the wave is propagating and $|\mathrm{Im}(k_y)|$ is the spatial attenuation. Therefore, the wave dispersion characteristics can be determined on a given-frequency form, enabling the computation of forced responses from a wave-based method. An example of the results obtained by a given-($\omega, k_y$) formulation is described in Fig.~\ref{fig:wave_directivity} for a thick-layered plate. Since $\omega$ is given in this formulation, the group velocities can be determined to evaluate the waves directivity in 2D waveguides.	
	
	Yet, the latter formulation requires a dynamic condensation of the inner DOFs for each frequency step. It may lead to extensive computation time when the periodic cell is investigated in higher frequencies, or contains a complex geometry and needs refined FEM. Besides, the condensed eigenproblem equation~(\ref{eq2.3akrout}) still involves numerous DOFs and remains a major drawback for analysing structurally advanced periodic waveguides. In the following section a hybrid formulation is proposed which combines the use of a modal reduction approach for the dynamic condensation with a wave shape frequency-interpolation.						

\section{Proposed model order reduction strategy}\label{sec3}
It should be noticed that the proposed formulation is a two-scale approach \cite{Langley1997}, using hybrid CMS formulation in a wave approach context. The CMS, described in Section~\ref{sec3.1} is performed at the cell's scale (mesoscopic scale), while a frequency-interpolation method based on propagating waves determination is detailed in Sections \ref{sec3.2}--\ref{sec3.4}. As a convention, the symbol $\boldsymbol{\Phi}$ is used for modes and $\boldsymbol{\Psi}$ is used for waves.

	\subsection{A Craig--Bampton reduction for the cell's inner DOFs}\label{sec3.1}
	The CMS procedure provides a significant reduction of the number of inner DOFs involved by replacing physical displacements by the cell's local modes. To this aim, the displacement vector $\mathbf{q}$ defined by Eq.~(\ref{eqgoverning}) is partitioned into the inner cell's displacements $\mathbf{q}_I$, and the boundary displacements:
	\begin{equation}
	\mathbf{q}_b = \left[\mathbf{q}_B^T \quad \mathbf{q}_T^T \quad \mathbf{q}_L^T \quad \mathbf{q}_R^T \quad \mathbf{q}_1^T \quad \mathbf{q}_2^T \quad \mathbf{q}_3^T \quad \mathbf{q}_4^T \right]^T
	\end{equation}		
	Since there are no external forces applied on the cell's inner DOFs, $\mathbf{f}_I = \mathbf{0}$, Eq.~(\ref{eqgoverning}) can then be written:
	\begin{equation}
	\left(
	\left[ \begin{array}{cc}
	\mathbf{K}_{\tt b} & \mathbf{K}_{\tt bI} \\
	\mathbf{K}_{\tt Ib} & \mathbf{K}_{\tt I} 
	\end{array} \right] - \omega^2 
	\left[ \begin{array}{cc}
	\mathbf{M}_{\tt b} & \mathbf{M}_{\tt bI} \\
	\mathbf{M}_{\tt Ib} & \mathbf{M}_{\tt I} 
	\end{array} \right] \right) 
	\left( \begin{array}{c}
	\mathbf{q}_b \\
	\mathbf{q}_I
	\end{array} \right) = 
	\left( \begin{array}{c}
	\mathbf{f}_b \\
	\mathbf{0}
	\end{array} \right)
	\end{equation}
	Therefore, a reduced set of coordinates is required to replace the physical DOFs $\mathbf{q}_I$. This reduced Ritz basis involves vectors associated with static boundary modes $\boldsymbol{\Phi}_{\tt b}$ and selected component modes $\boldsymbol{\Phi}_{\tt C}$ so that the displacements can be written:
	\begin{equation}\label{craig}
	\left( \begin{array}{c}
	\mathbf{q}_b \\ 
	\mathbf{q}_I
	\end{array} \right)
	 = \left[ \begin{array}{cc} 
	\mathbf{I} & \mathbf{0} \\ 
	\boldsymbol{\Phi}_{\tt b} 	& \boldsymbol{\Phi}_{\tt C}
	\end{array} \right] 
	\left( \begin{array}{c}
	\mathbf{q}_b \\ 
	\mathbf{P}_C
	\end{array} \right)
	 = \mathbf{B} \left( \begin{array}{c}
	\mathbf{q}_b \\ 
	\mathbf{P}_C
	\end{array} \right)
	\end{equation}
	where the vectors $\boldsymbol{\Phi}_{\tt b}$ are obtained by static condensation
	\begin{equation}\label{static}
	\boldsymbol{\Phi}_{\tt b} = -\mathbf{K}_{\tt I}^{-1} \mathbf{K}_{\tt Ib}
	\end{equation}
	the component modes $\boldsymbol{\Phi}_{\tt C}$ are selected among the eigenvectors $\mathbf{\Phi}_{\tt I}$ of Eq.~(\ref{component})
	\begin{equation}\label{component}
	\left( \mathbf{K}_{\tt I} - \omega^2 \mathbf{M}_{\tt I} \right) \mathbf{\Phi}_{\tt I} = \mathbf{0}
	\end{equation}
	and the inner DOFs displacements are replaced with a reduced set of modal participations $\mathbf{P}_C$. In Craig--Bampton procedure, modal selection is based on the lower resonance frequencies. This method can be extended in a wave approach context since the aim here is to capture the local deformed shape of the periodic cell. The underlying hypothesis is that when a wave propagates through the periodic waveguide, the displacements inside a unit-cell can be expanded on a subset of stationary modes. Therefore, one can choose \cite{Craig1968} modes into the frequency range $\left[0 , 2\, \omega_{\tt max}\right]$, where $\omega_{\tt max}$ is the maximum frequency of interest for the wave dispersion analysis. Then, the stiffness and mass matrices can be written in the reduced set of coordinates using the projection matrix $\mathbf{B}_\varPhi$ defined by Eq.~(\ref{craig}):
	\begin{equation}\label{CMS}
	\mathbf{\tilde K}= \mathbf{B}_\varPhi^T  \mathbf{K} \mathbf{B}_\varPhi
	, \quad
	\mathbf{\tilde M}= \mathbf{B}_\varPhi^T  \mathbf{M} \mathbf{B}_\varPhi
	\end{equation}
It should be noted that free modes could also be used for the CMS procedure. Yet, they do not improve the reduced model accuracy and require additional projection of the boundary degrees of freedom.
	As mentioned in Section~\ref{2.3}, the reduced matrices can be used to perform dynamic condensation in order to obtain a spectral problem of size $ \left[ \mathbf{q}_1^T \quad \mathbf{q}_L^T \quad \mathbf{q}_B^T \right]^T$. However the size of this state vector remains a major drawback when a large frequency band is considered. A wave-based reduced formulation for broadband analysis of 2D homogeneous or periodic waveguides is developed in the following.
	
	\subsection{Determination of a reduced frequency sample for the wave shape interpolation}\label{sec3.2}
	
	The principle of the model order reduction (ROM) strategy proposed here is to reduce Eq.~(\ref{eq2.3akrout}) using a subset of the propagating wave solutions. The main assumption is that a waveguide cross-sectional deformed shape for a given wave type at a given frequency can be used to describe the motion of the periodic cell for the same wave in a large frequency range around this given frequency. It will be considered further in this paper for 2D periodic problems. In order to identify a suitable reduced subset $\tilde{\varOmega}$ of the frequency range $\varOmega$, one can consider the specific behaviour of propagating waves in periodic waveguides. Indeed, in such structures the real wavenumbers remain between $k = 0$ and $k = \pi/d$ (see \cite{MeadJSV1979}). The wave shapes are then retained at the mid-aliasing frequencies, defined by the wavenumber: $k = \pi/(2d)$. Therefore, the frequency subset can be determined by using given $(k_x, k_y)$ formulation. From the reduced matrices defined by Eq.~(\ref{CMS}), the dispersion relation defined by Eq.~(\ref{eq2.2eig}) can be written using reduced modal participations:
	\begin{equation}\label{spectralkxky}
	\mathbf{\Lambda}_L' (\lambda_x, \lambda_y) \left( \mathbf{\tilde{K}} - \omega^2 \mathbf{\tilde{M}}\right) \mathbf{\Lambda}_L'^{H}(\lambda_x, \lambda_y)
	\left( \begin{array}{c}
	\mathbf{q}_1 \\ 
	\mathbf{q}_L \\ 
	\mathbf{q}_B \\ 
	\mathbf{P}_C
	\end{array} \right) = \mathbf{0}
	\end{equation}
	where $\mathbf{\Lambda}_L'$ is the definite positive Hermitian matrix detailed in \ref{annexe1}. 
	In this given $(k_x, k_y)$ formulation, the propagating solutions are associated with the real eigenvalues of Eq.~(\ref{spectralkxky}). These solutions form the frequency subset $\tilde{\varOmega}$ used in the following to span the reduced wave basis.
	
	\subsection{Resolution of the spectral problem on the frequency subset $\Delta\omega$}\label{sec3.3}
	
	Considering matrices $\mathbf{\tilde{K}}$ and $\mathbf{\tilde{M}}$ defined in Section~\ref{sec3.1} and the given frequency $\tilde{\omega} \in \tilde{\varOmega}$, the reduced dynamic stiffness defined as $\mathbf{\tilde{D}} = \mathbf{\tilde{K}} - \tilde{\omega}^2 \mathbf{\tilde{M}}$ can be ordered as follows:
	\begin{equation}
	\left(\begin{array}{cc}
	\mathbf{\tilde{D}}_{\tt b} & \mathbf{\tilde{D}}_{\tt bI} \\
	\mathbf{\tilde{D}}_{\tt Ib} & \mathbf{\tilde{D}}_{\tt I} 
	\end{array}\right)
	\left(\begin{array}{c}
	\mathbf{q}_b \\
	\mathbf{P}_C  
	\end{array}\right) =
	\left(\begin{array}{c}
	\mathbf{f}_b \\
	\mathbf{0} 
	\end{array}\right)
	\end{equation}
	Then, the modal participations $\mathbf{P}_{\tt C}$ can be condensed on the cell's boundaries so that:
	\begin{equation}\label{condens}
	\mathbf{D} = \mathbf{\tilde{D}}_{\tt b} - \mathbf{\tilde{D}}_{\tt bI}\mathbf{\tilde{D}}_{\tt I}^{-1}\mathbf{\tilde{D}}_{\tt Ib}
	\end{equation}
	The dynamic stiffness can then be ordered as in Eq.~(\ref{Dcond}), so that the spectral problem can be written in terms of this condensed matrix:
	\begin{equation}\label{spectralcond}
	\mathbf{\Lambda}_L (\lambda_x, \lambda_y) \, \mathbf{D}(\omega) \, \mathbf{\Lambda}_R(\lambda_x, \lambda_y)
	\left( \begin{array}{c}
	\mathbf{q}_1 \\ 
	\mathbf{q}_L \\ 
	\mathbf{q}_B
	\end{array} \right) = \mathbf{0}
	\end{equation}
	In this case the problem is not linear since $\lambda_x$ is unknown and $(\omega, \lambda_y)$ are given. Therefore, a quadratic eigenproblem in $\lambda_x$ can be formed using Eqs.~(\ref{spectralcond},\ref{Dcond}):
	\begin{equation}\label{eq:polynome}
	\left(	\lambda _x\left[ \begin{array}{ccc}
	\mathbf{A}_{\tt 11} & \mathbf{A}_{\tt 1L} & \mathbf{A}_{\tt 1B} \\
	\mathbf{A}_{\tt L1} & \mathbf{A}_{\tt LL} & \mathbf{A}_{\tt LB} \\
	\mathbf{A}_{\tt B1} & \mathbf{A}_{\tt BL} & \mathbf{A}_{\tt BB}
	\end{array} \right] +
	\left[ \begin{array}{ccc}
	\mathbf{B}_{\tt 11} & \mathbf{B}_{\tt 1L} & \mathbf{B}_{\tt 1B} \\
	\mathbf{B}_{\tt L1} & \mathbf{B}_{\tt LL} & \mathbf{B}_{\tt LB} \\
	\mathbf{B}_{\tt B1} & \mathbf{B}_{\tt BL} & \mathbf{B}_{\tt BB}
	\end{array} \right] +
	\frac{1}{\lambda _x} \left[ \begin{array}{ccc}
	\mathbf{C}_{\tt 11} & \mathbf{C}_{\tt 1L} & \mathbf{C}_{\tt 1B} \\
	\mathbf{C}_{\tt L1} & \mathbf{C}_{\tt LL} & \mathbf{C}_{\tt LB} \\
	\mathbf{C}_{\tt B1} & \mathbf{C}_{\tt BL} & \mathbf{C}_{\tt BB}
	\end{array} \right] \right) 
	\left( \begin{array}{c}
	\mathbf{q}_1 \\ 
	\mathbf{q}_L \\ 
	\mathbf{q}_B
	\end{array} \right) = \mathbf{0}
	\end{equation}
	where the coefficients $\mathbf{A}_{\tt ij}$, $\mathbf{B}_{\tt ij}$ and $\mathbf{C}_{\tt ij}$ are detailed in \ref{annexe2}. Then denoting these matrices as $\mathbf{A}$, $\mathbf{B}$, $\mathbf{C}$, the quadratic eigenproblem equation~(\ref{eq:polynome}) can be solved using the following linearization:
	\begin{equation}\label{linearised}
	\left[ \begin{array}{cc}
	-\mathbf{C} & \mathbf{0}  \\
	\mathbf{0} & \mathbf{I}  
	\end{array} \right] 
	\left( \begin{array}{c}
	\mathbf{q} \\ 
	\lambda_x \mathbf{q}
	\end{array} \right)
	=	\lambda_x	\left[ \begin{array}{cc}
	\mathbf{B} & \mathbf{A}  \\
	\mathbf{I} & \mathbf{0}  
	\end{array} \right] 
	\left( \begin{array}{c}
	\mathbf{q} \\ 
	\lambda_x \mathbf{q}
	\end{array} \right)
	\end{equation}
	For the sake of clarity, the spectral problem equation~(\ref{eq:polynome}) will be denoted as $\mathbf{S}\left(\lambda _x,\lambda _y, \omega \right) \varPsi = \mathbf{0}$, where the solution $\varPsi$ is the wave shape on the condensed DOFs defined as:
	\begin{equation}
	\varPsi = \left( \begin{array}{c}
	\mathbf{q}_1 \\ 
	\mathbf{q}_L \\ 
	\mathbf{q}_B
	\end{array} \right)
	\end{equation}
	The solutions of Eq.~(\ref{eq:polynome}) are computed on the mid-aliasing frequency subset $\tilde{\varOmega}$ defined in Section~\ref{sec3.2}. Accordingly, an appropriate reduced basis for the ROM can be spanned using solutions of the following system of equations:
	\begin{equation}\label{systemeig}
	\begin{cases}
	\mathbf{S}\left(\lambda _x,\lambda _y, \tilde{\omega}_{\tt 1} \right) \varPsi = \mathbf{0}	\\
	\vdots \\
	\mathbf{S}\left(\lambda _x,\lambda _y, \tilde{\omega}_{\tt j} \right) \varPsi = \mathbf{0} & ; \quad \tilde{\omega}_{\tt j} \in \tilde{\varOmega} \\
	\vdots \\
	\mathbf{S}\left(\lambda _x,\lambda _y, \tilde{\omega}_{\tt n} \right) \varPsi = \mathbf{0}	\\
	\end{cases}
	\end{equation}
	where $n$ is the number of elements in the subset $\tilde{\varOmega}$. For each given $\tilde{\omega}_{\tt j}$, discrete values of $k_y$ are chosen in the range $[-\pi/d_y,  \pi/d_y]$. The values of $\lambda_y$ are derived from these choices of $k_y$, with $\lambda_y = e^{-jd_y k_y}$. It is noteworthy that the CPU time required to solve the $n$ eigenproblems equation~(\ref{systemeig}) is considerably lower than for solving Eq.~(\ref{eq:polynome}) on the whole frequency domain $\varOmega$. The reduced basis can be built using these eigenvectors. However, each line of Eq.~(\ref{systemeig}) yields $2\times N$ solutions, where $N$ is the size of $\varPsi = \left[\mathbf{q}^T_1 , \mathbf{q}^T_L , \mathbf{q}^T_B \right]^T$, corresponding to the number of DOFs in the state vector. Retaining all of the $2\times N$ solutions $\varPsi$ at the $n$ frequencies would lead to an extensive wave shape basis, while the only solutions of interest are associated with propagating waves. This issue is discussed below.
	
	\subsection{Building a reduced wave shape basis}\label{sec3.4}
	
	The idea is to select propagating waves using their dispersion characteristics in order to build a reduced wave basis. A complex wavenumber $k_x = j\ln (\lambda_x)$ is associated with each eigenvector $\varPsi$, and its imaginary part $\mathrm{Im}(k)$ is the wave attenuation. The wave is evanescent when the wavenumber is purely imaginary, while real positive wavenumbers are associated with the least decaying positive-going waves. Yet, damped structures can produce highly decaying waves and the determination of propagating solutions can be delicate. In this study, a propagating wave is described by the following conditions:
	\begin{equation}\label{kfilter}
	\begin{cases}
	0 \leq \mathrm{Re}(k_x) \leq \pi/d_x & \text{Positive-going waves} \\
	\mathrm{Im}(k_x) \leq 0 & \text{Physical solutions} \\
	|\mathrm{Im}(k_x)| \leq \alpha\, \mathrm{Re}(k_x) & \text{Least-decaying waves,} \quad \alpha\simeq 0.01
	\end{cases}
	\end{equation}
	We denote $\varPsi^{(j)}_i$ the eigenvector associated with the i-th propagating eigenvector at the frequency $\tilde{\omega}_j$. Therefore, the solution subspace can be written:
	\begin{equation}
	\mathbf{\Psi} = \lbrace \varPsi^{(j)}_1 , \ldots, \varPsi^{(j)}_{p(j)} \rbrace_{1\leq j\leq n}
	\end{equation}
	where $p(j)$ stands for the number of propagating waves retained at frequency $\tilde{\omega}_j$. We denote $p = \sum_{i=1}^n p(i)$ the total number of vectors in $\mathbf{\Psi}$. For the sake of clarity the j-th column of matrix $\mathbf{\Psi}$ is a vector, denoted $\mathbf{\Psi}(j)$. In practice, this set may involve redundant vectors and lead to an unnecessary large wave basis when a wave shape remains the same at several frequencies in $\tilde{\varOmega}$. Considering modal assurance criteria (MAC) as an indicator of correlation between two complex eigenvectors, the orthogonality relation $\perp_\varepsilon$ can be written:
	\begin{equation}\label{orthog}
	\psi_1 \perp_\varepsilon \psi_2 \quad \Leftrightarrow \quad \left| \frac{(\psi_1^T \bar{\psi}_2)\, (\psi_2^T \bar{\psi}_1)}
	{(\psi_1^T \bar{\psi}_1)\, (\psi^T_2 \bar{\psi}_2)}\right| \leq \varepsilon
	\end{equation}
	The criterion $\varepsilon \in \left[ 0\, , \, 1 \right]$ is chosen close to 1 and describes the tolerance for the orthogonality relation. This definition is extended to define the orthogonality between a vector and a set $\psi_1 \perp_\varepsilon \left\lbrace \mathbf{\psi}\right\rbrace$, when $\psi_1$ is orthogonal to all the vectors in $\left\lbrace \psi\right\rbrace$. An optimization procedure is required to reduce the number of vectors in the wave basis. In order to create this subset $\mathbf{\tilde{\Psi}}$ of $\mathbf{\Psi}$ by elimination of correlated wave shapes and build a reduced wave basis providing good numerical stability, we propose the algorithm described in Eq.~(\ref{iterative}).
	
	First, the vector subset $\mathbf{\tilde{\Psi}}$ is built by recurrence using $\perp_\varepsilon$ orthogonality. Note that the choice of $\varepsilon$ influences the size of the reduced basis and the ROM efficiency. Then, a simple way to improve numerical stability is to normalize the remaining vectors and dissociate their real and imaginary parts. Thus, the wave basis $\mathbf{\Gamma}$ is purely real and well-conditioned.
	\begin{equation}\label{iterative}
	\begin{cases}
	\quad \texttt{Define} \quad \mathbf{\tilde{\Psi}}(1) = \mathbf{\Psi}(1) \\[2mm]
	\quad \texttt{for}\quad j=1  \quad \texttt{to} \quad j=p-1 \\[2mm]
		\quad \begin{cases}
		\quad \texttt{if} \quad \mathbf{\Psi}(j+1) \perp_\varepsilon \left\lbrace\mathbf{\tilde{\Psi}}\right\rbrace \\[2mm]
		\quad \quad \texttt{do} \quad \left\lbrace \mathbf{\tilde{\Psi}} \right\rbrace = \left[\,\left\lbrace\mathbf{\tilde{\Psi}}\right\rbrace, \mathbf{\Psi}(j+1)\,\right] \\[2mm]
		\quad \texttt{end if}
		\end{cases}\\[5mm]
	\quad \texttt{end for}\\
	\quad \texttt{define the normalized} \quad \varGamma_i = \dfrac{\mathbf{\tilde{\Psi}}(i)}{\|\mathbf{\Psi}(i)\|}\\[2mm]
	\quad \texttt{dissociate complex vectors:} \quad \varGamma_i \quad \texttt{into:} \quad \left[ \mathrm{Re}(\varGamma_i) \quad \mathrm{Im}(\varGamma_i)\right]\\[2mm]
	\quad \texttt{build the wave basis} \quad \mathbf{\Gamma} = \left[\varGamma_1 , \ldots, \varGamma_r\right]
	\end{cases}
	\end{equation}
	The state vector can be written in terms of wave participations:
	\begin{equation}
	\left( \begin{array}{c}
	\mathbf{q}_1\\
	\mathbf{q}_L\\
	\mathbf{q}_B
	\end{array} \right) = 	
	\left[ \begin{array}{ccc}
	\mathbf{\Gamma}^{(1)} & \mathbf{0} & \mathbf{0} \\[1mm]
	\mathbf{0} & \mathbf{\Gamma}^{(L)} & \mathbf{0} \\[1mm]
	\mathbf{0} & \mathbf{0} & \mathbf{\Gamma}^{(B)}
	\end{array}	\right] 
	\left( \begin{array}{c}
	\mathbf{w}_1 \\[1mm]
	\mathbf{w}_L \\[1mm]
	\mathbf{w}_B
	\end{array}	 \right) 
	\end{equation}
	where $\mathbf{\Gamma}^{(1)}$, $\mathbf{\Gamma}^{(L)}$ and $\mathbf{\Gamma}^{(B)}$ are associated to the displacements of the \textit{corner}, \textit{left} and \textit{bottom} edges of the periodic cell, while $\mathbf{w}_1$, $\mathbf{w}_L$, $\mathbf{w}_B$ are vectors of size $r$, containing the reduced wave participations. The global displacements vector in the periodic cell, involving modal participations as described in Section~\ref{sec3.1}, can be expanded on the wave basis as:
	\begin{equation}\label{expandedw}
	\left( \begin{array}{c}
	\mathbf{q}_1 \\	\mathbf{q}_2 \\	\mathbf{q}_3 \\	\mathbf{q}_4 \\
	\mathbf{q}_L \\	\mathbf{q}_B \\	\mathbf{q}_R \\	\mathbf{q}_T
	\end{array}\right)  = 
	\tilde{\mathbf{B}}_\varPsi \mathbf{w} = 
	\left[ \begin{array}{cccccccc}
	\mathbf{\Gamma}^{(1)} & \\
	 & \mathbf{\Gamma}^{(2)} & & & & & \mathbf{0} & \\
	 & & \mathbf{\Gamma}^{(3)} & \\
	 & & & \mathbf{\Gamma}^{(4)} & \\
	 & & & & \mathbf{\Gamma}^{(L)} & \\
	 & & & & & \mathbf{\Gamma}^{(B)} & \\
	 & \mathbf{0} & & & & & \mathbf{\Gamma}^{(R)} & \\
	 & & & & & & & \mathbf{\Gamma}^{(T)}
	\end{array} \right]  \left( \begin{array}{c}
	\mathbf{w}_1 \\	\mathbf{w}_2 \\	\mathbf{w}_3 \\	\mathbf{w}_4 \\
	\mathbf{w}_L \\	\mathbf{w}_B \\	\mathbf{w}_R \\	\mathbf{w}_T
	\end{array}\right)
	\end{equation}
	where $\mathbf{\Gamma}^{(1)} = \mathbf{\Gamma}^{(2)} = \mathbf{\Gamma}^{(3)} = \mathbf{\Gamma}^{(4)}$, $\mathbf{\Gamma}^{(L)} = \mathbf{\Gamma}^{(R)}$ and $\mathbf{\Gamma}^{(B)} = \mathbf{\Gamma}^{(T)}$. It is noteworthy that wave participations are defined separately for each of the periodic cell's components. Geometrically, it means that three different shape bases are defined for the corners and both pairs of opposite edges, and each of these components is provided with independent coefficients. Since the coefficients in \ref{annexe2} are linear with respect to matrix $\mathbf{D}$, one can build reduced dynamic stiffness $\tilde{\mathbf{B}}^T_\varPsi \mathbf{D} \tilde{\mathbf{B}}_\varPsi$. Then denoting:
	\begin{equation}\label{Bpsi}
	\mathbf{B}_\varPsi = \left[ \begin{array}{cc}
	\tilde{\mathbf{B}}_\varPsi & \mathbf{0} \\
	\mathbf{0} & \mathbf{I}
	\end{array}\right] 
	\end{equation}
	and using Eq.~(\ref{CMS}), one can write reduced mass and stiffness matrices leading to a reduced CPU computation for the dynamic condensation equation~(\ref{condens}) and the spectral eigenproblem equation~(\ref{eq:polynome}):
	\begin{equation}\label{Kpsi}
	\tilde{\mathbf{K}}_\varPsi =  \mathbf{B}_\varPsi^T \mathbf{B}_\varPhi^T \, \mathbf{K} \, \mathbf{B}_\varPhi \mathbf{B}_\varPsi 
	, \quad
	\tilde{\mathbf{M}}_\varPsi =  \mathbf{B}_\varPsi^T \mathbf{B}_\varPhi^T \, \mathbf{M} \, \mathbf{B}_\varPhi \mathbf{B}_\varPsi 
	\end{equation}
	
	To summarize, a modal basis is defined using Eqs.~(\ref{craig})--(\ref{CMS}) in order to replace the inner DOFs of the periodic cell with a reduced set of modal participations so that the dynamic condensation equation~(\ref{condens}) can be easily done on the frequency range $\varOmega$. Then, the spectral problem equation~(\ref{eq:polynome}) is solved on a small frequency subset $\tilde{\varOmega}$, defined using Eq.~(\ref{spectralkxky}). The solutions of system equation~(\ref{systemeig}) associated with positive-going propagating waves are selected using the conditions equation~(\ref{kfilter}), and a reduced set of eigenvectors is built using the algorithm provided in Eq.~(\ref{iterative}). Therefore, the wave participations are defined using Eq.~(\ref{expandedw}) and reduced mass and stiffness matrices are provided using Eqs.~(\ref{Bpsi}) and (\ref{Kpsi}). Finally, a reduced size spectral problem equation~(\ref{eq:polynome}) can be solved on the whole frequency range $\varOmega$, while the physical solutions can be derived using Eqs.~(\ref{craig}), (\ref{expandedw}).
	
\section{Numerical results: orthogonally stiffened panels}\label{sec4}
	\subsection{Problem description}
	
\begin{figure}[!ht]
  \centering
  \includegraphics[width=0.75\linewidth]{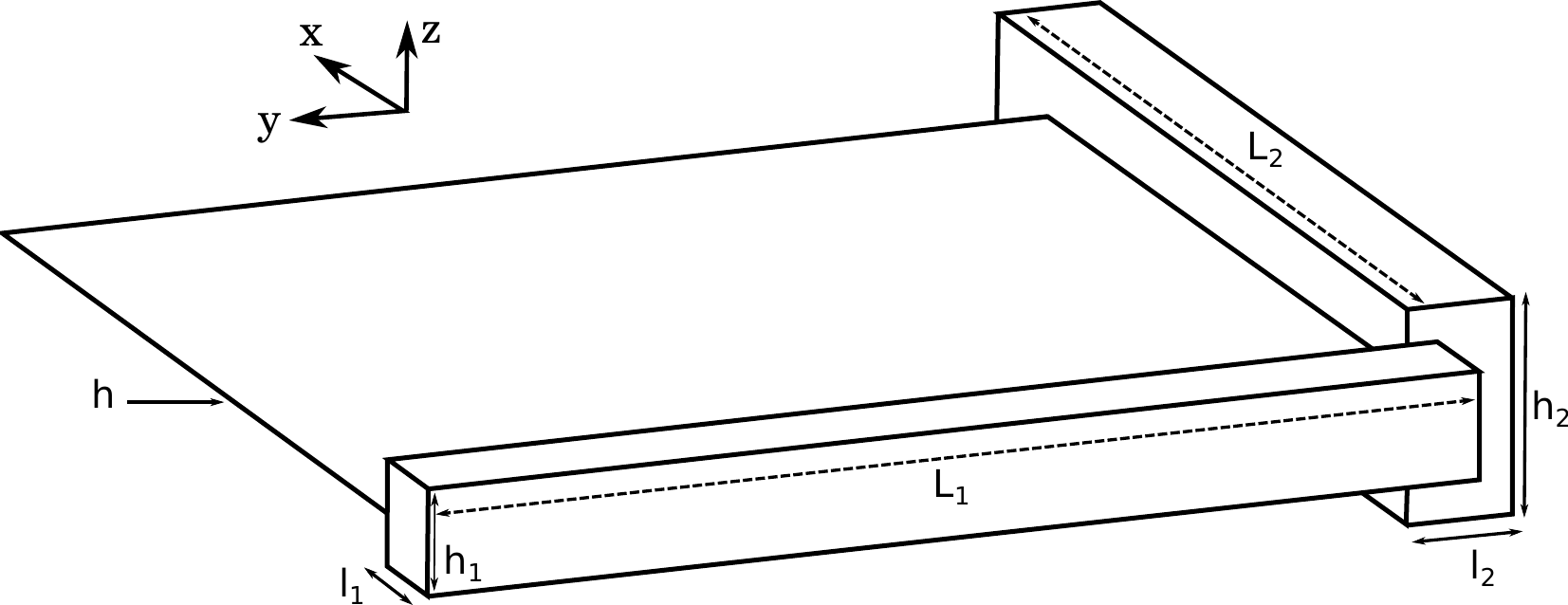}
  \caption{Illustration of the stiffened panel's periodic cell.}
	   \label{stiff_cell}
	\end{figure}

	In this section, the formulation developed in the paper is used to provide wave dispersion characteristics (wavenumbers, group velocities, wave modes) in bidirectionally stiffened panels. The wave propagation in such structures is extensively studied in the literature \cite{Orrenius1996, Legault2011}. The mode selection and wave-based interpolation parameters introduced in Section~\ref{sec3} are used here to study the stability and accuracy of the proposed method. A flat panel reinforced by stiffeners with rectangular cross-sections is considered in this example. The unit-cell of the periodic plate is shown in Fig.~\ref{stiff_cell}. The skin is an aluminium plate of thickness $h=0.8$\,mm and the stiffeners are rectangular aluminium beams of width $l_1 = l_2 = 10$\,mm and height $h_1=20$\,mm and  $h_2=10$\,mm. A structural loss factor $\eta =$ 0.5 percent is introduced and the spacing between the stiffeners is $L_1=0.08$\,m in the y-direction and $L_2=0.12$\,m in the x-direction. The FE model built using ANSYS 14.0 Software involves 600 4-nodes SHELL63 (DKT) elements for the skins, and 50 rectangular BEAM44 elements with cubic interpolation for the stiffeners. Therefore, the periodic cell has a total of 3906 DOFs, involving 3306 DOFs inside the periodic cell and 600 DOFs on the cell's boundaries. Since the stiffened panel exhibits two localized resonances at 840 Hz and 2040 Hz, the frequency range studied in this application is $\left[ 0 - 2500 \right] $\,Hz.
	
	\subsection{Reduced WFEM solution}
	
\begin{figure}[!ht]
  \centering
  \includegraphics[width=0.45\linewidth]{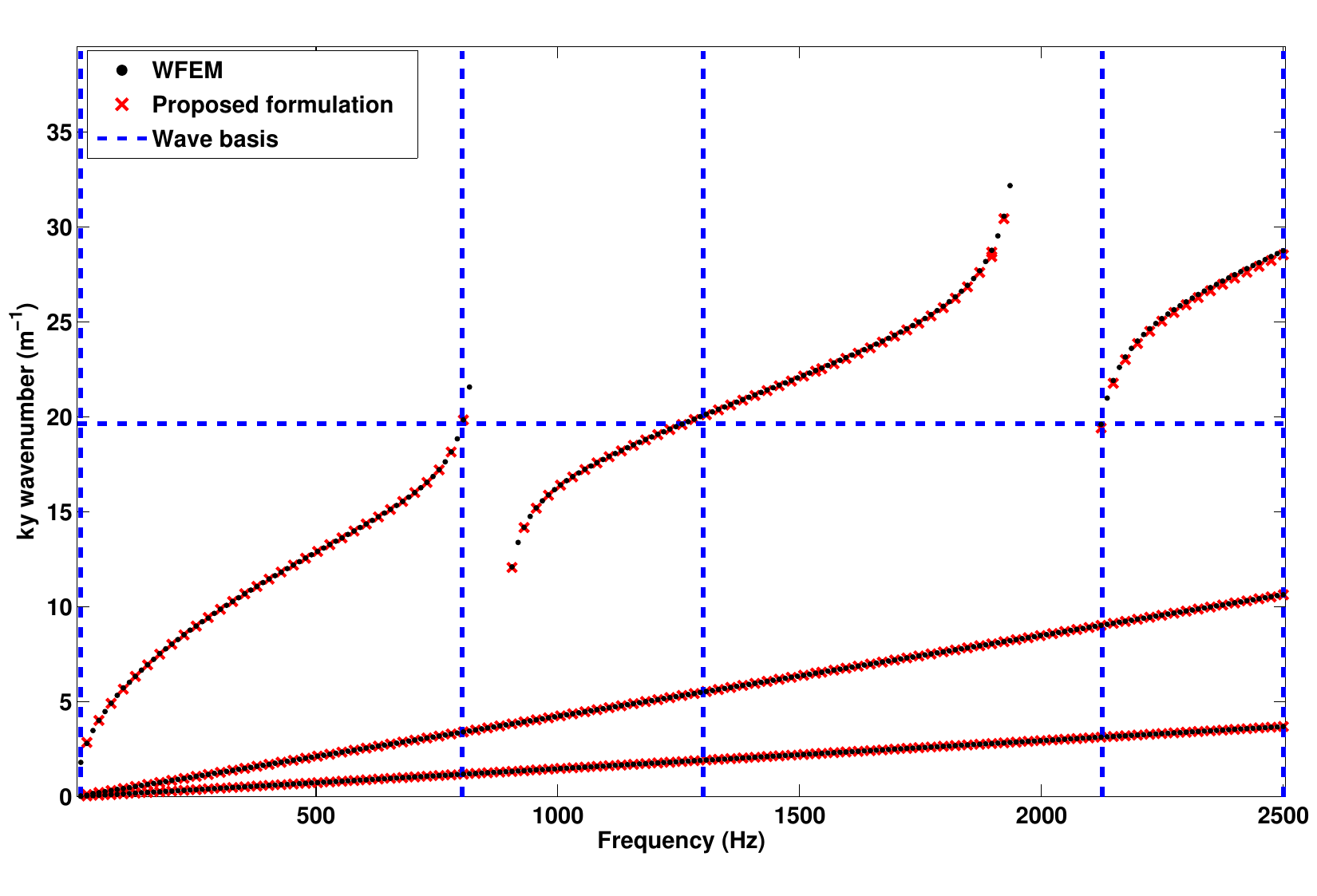}
  \caption{Real wavenumbers in the $y$-direction (for $k_x = 0$) using the standard WFE method and the proposed CMS-wave interpolation strategy.}
	   \label{RCWFEM}
	\end{figure}

	The proposed reduction strategy is applied to the model described above and compared to the standard formulation of the WFEM. Five stationary modes are employed to describe the inner cell's displacements, according to the CMS procedure described Section~\ref{sec3.1} and 15 waves are determined using the procedure equation~(\ref{iterative}) to build the reduced solution subspace. Real wavenumbers associated with the propagating waves in the x-direction are presented in Fig.~\ref{RCWFEM} using the standard WFEM and the proposed formulation. Vertical lines, where the full eigenproblems are solved denote the mid-aliasing frequencies. Note that all of the intersections between the propagating waves and the mid-aliasing frequency are correctly predicted. A very good correlation is observed between the standard WFEM formulation and the reduced model, involving modal condensation and wave-based interpolation. The maximal error on the wavenumbers is $err = 0.8$ percent for the flexural wave. The modal basis is built using a subset of stationary modes between $0$ and $2500$\,Hz, and the wave basis is computed using a correlation criterion $\varepsilon = 0.99$ described in Eq.~(\ref{orthog}), and provides a reduction of the quadratic eigenproblem from 600 boundary DOF's to 120 wave participations. 
	
A significant reduction of the computational cost is observed for the proposed method. CPU time using Matlab 7.14 running on Linux with two AMD 64$\times$2 dual core 6000+ processors are given in Table\,\ref{CPU}. The CMS procedure requires an additional modal analysis, but provides an important reduction (95.5 percent) of the time required for the dynamic condensation. On the other hand, the wave based reduction strategy also provides a considerable reduction (95.7 percent) of the time required to solve the quadratic eigenproblems, in comparison with the standard formulation involving the edges unit-cell nodal displacements. Therefore, the proposed formulation provides a total of 80 percent reduction of the computational cost. Noteworthy that a higher frequency sampling will increase the wave-based reduction efficiency. 
	\begin{table}[!ht]
	\begin{center} 
	\begin{tabular}{|lcccc|}
	\hline
	Time per iteration (s) 					    & \bf{WFEM} & CMS-WFEM   & Interp-WFEM & \bf{Proposed method} \\
	\hline \\
	\textit{Modal analysis ($\times$1)} 	    & --	         &  258           & --	      & 258    \\
	\textit{Dynamic condens. ($\times$500)} & \textbf{1.85}      &  0.082         & 1.85              & \textbf{0.082} \\
	\textit{Building wave basis ($\times$1)}    & --	         & 	--	  &  17 	          & 8.16   \\
	\textit{Spectral problem ($\times$500)}     & \textbf{1.55}      &  1.55          &  0.066            & \textbf{0.066} \\\\
	\hline
	\textbf{Total Time}	                        & \textbf{1700}      & 1074 (-37$\%$) & 975 (-43$\%$)     & \textbf{340 (-80$\%$)} \\
	\end{tabular}
	\caption{Stiffened plate: CPU time for the standard and reduced WFE formulations.\label{CPU}}
	\end{center}
	\end{table}
	
\begin{figure}[!ht]
  \centering
  \includegraphics[width=0.45\linewidth]{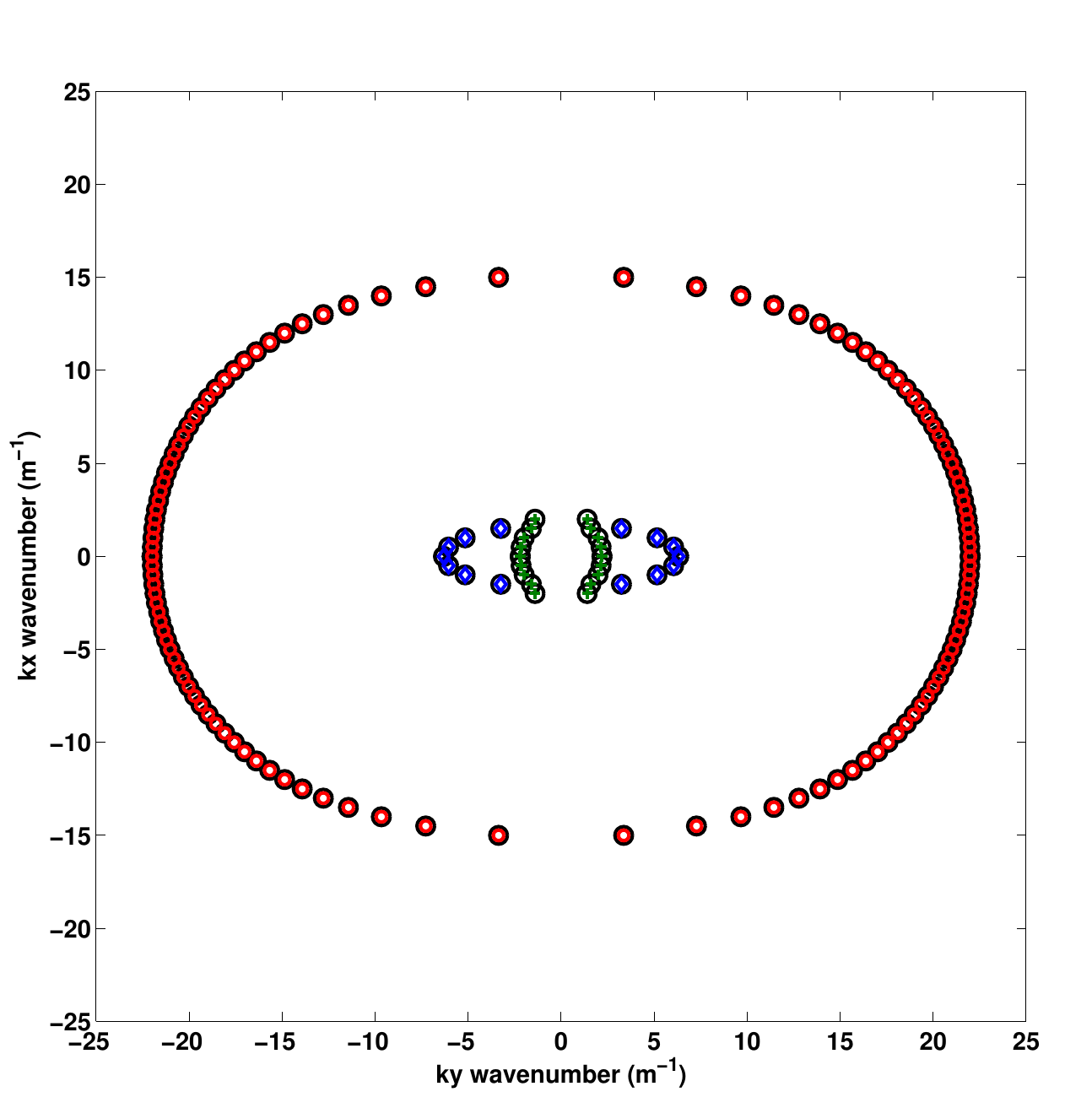}
  \caption{Comparison of the k-space solutions at 1500 Hz using standard WFEM and the proposed formulation (black `$\bigcirc$').}
	   \label{validation_kspace1}
	\end{figure}

\begin{figure}[!ht]
  \centering
  \includegraphics[width=0.45\linewidth]{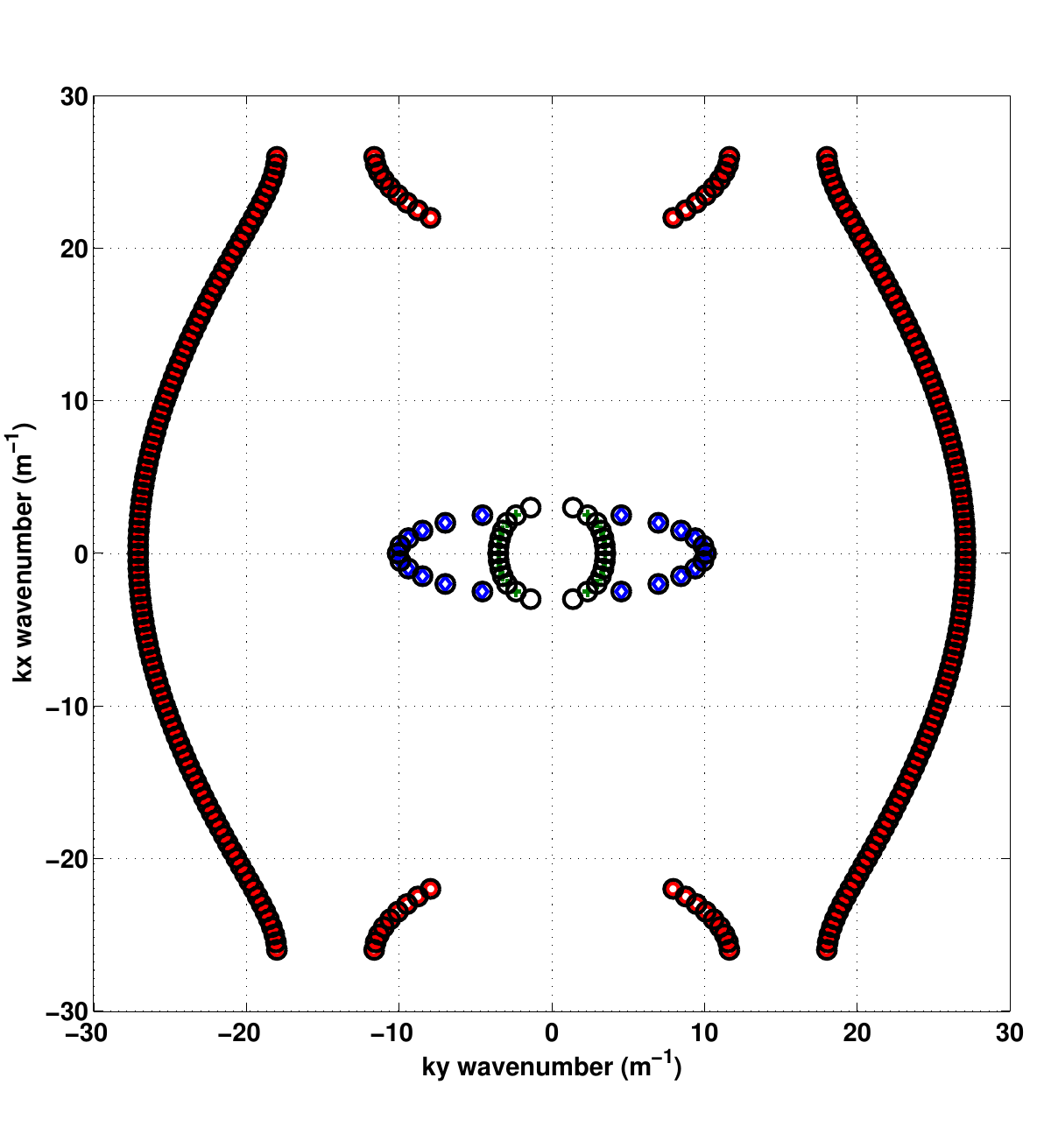}
  \caption{Comparison of the k-space solutions at 2400 Hz using standard WFEM and the proposed formulation (black `$\bigcirc$').}
	   \label{validation_kspace2}
	\end{figure}

	The aforementioned reduction strategy is then applied in the 2D k-space, and compared to the given - ($\omega,k_y$) standard WFE formulation, presented in Section~\ref{2.3}. The wavenumbers $k_x$ and $k_y$ are shown in Fig.~\ref{validation_kspace1} at 1500 Hz, and exhibit a very good correlation between the standard WFEM and the proposed method. The flexural ($\bigcirc$), shear ($\diamond$) and longitudinal waves ($+$) are accurately predicted by the reduced formulation ('\textbf{O}'). Besides, in Fig.~\ref{validation_kspace2} the localized resonances, associated with important variations of the directivity and appearance of angular bandgaps for the flexural wave, are predicted as well at 2400 Hz. At this frequency, the reduced model exhibits a good accuracy, although $k_x$ is given for all the values of the first Brillouin zone. Therefore, the k-space can be accurately determined in broadband frequency range and provide the required 2D wave dispersion characteristics of the periodic structure.

\subsubsection{Sensitivity of the CMS to the modal basis}
\begin{figure}[!ht]
  \centering
  \includegraphics[width=0.55\linewidth]{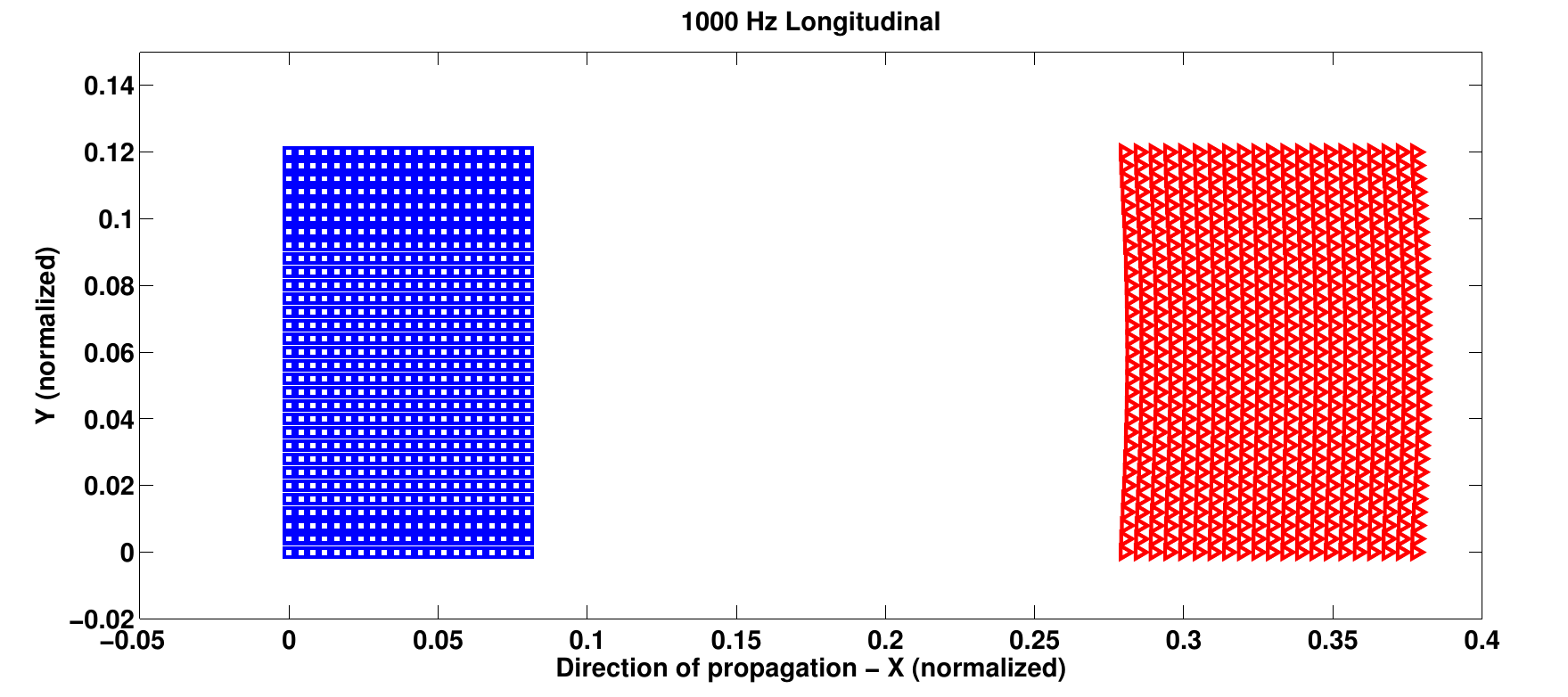}
  \caption{Deformed shape of the periodic cell for the longitudinal wave at 1000 Hz. Blue `$\square$': undeformed; red `$\triangleright$': deformed shape.}
	   \label{shape_long}
\end{figure}

\begin{figure}[!ht]
  \centering
  \includegraphics[width=0.35\linewidth]{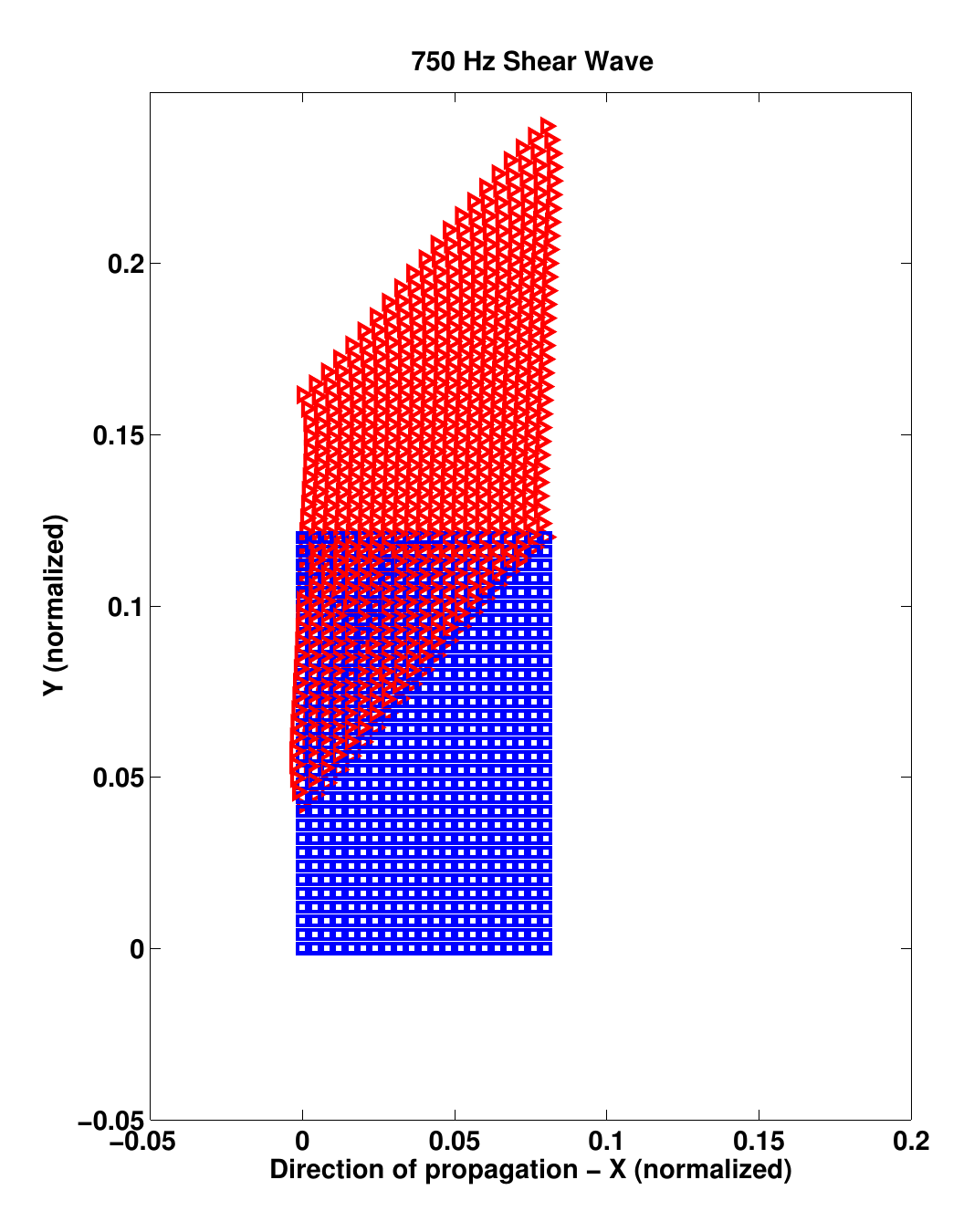}
  \caption{Deformed shape of the periodic cell for the shear wave at 750 Hz. Blue `$\square$': undeformed; red `$\triangleright$': deformed shape.}
	   \label{shape_shear}
\end{figure}

	The accuracy of the proposed ROM strategy relies essentially on two parameters: the number of modes retained for the CMS procedure and the number of wave shapes used to span the solution subspace. For the CMS procedure, the fixed-interface modes are usually retained between 0 and two times the maximal frequency. An insufficient number of modes is expected to produce significant errors in the determination of high-order wave shapes, since the displacement field inside the periodic cell can no longer be described. 
	\begin{figure}[!ht]
  \centering
  \includegraphics[width=0.75\linewidth]{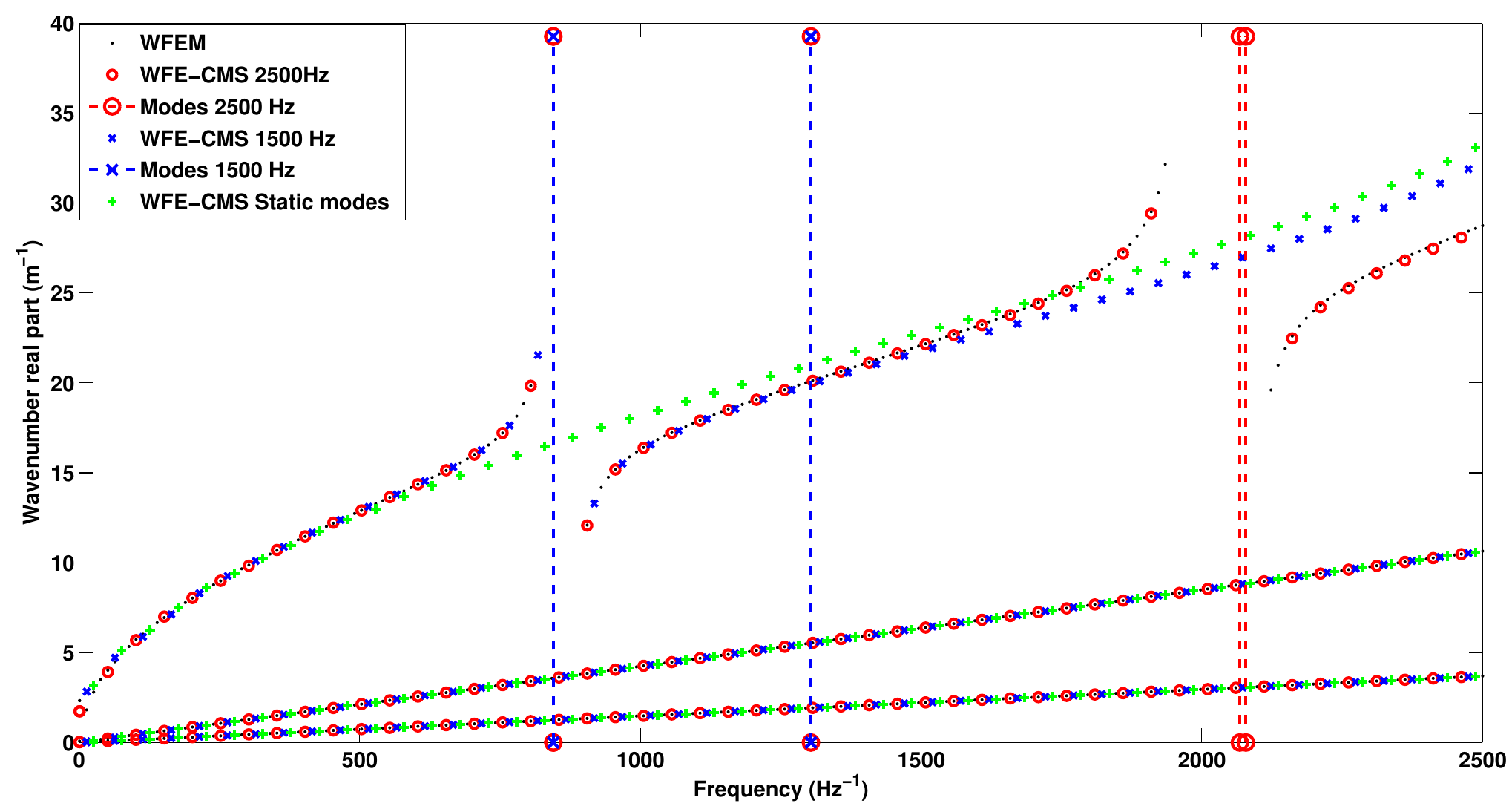}
  \caption{Influence of the modal basis used in the CMS procedure. Real part of wavenumbers in the y-direction ($k_x = 0$).}
	   \label{CMSparam}
	\end{figure}
	In Fig.~\ref{CMSparam}, the dispersion curves are compared using three different mode selection for the modal synthesis. The first basis (CMS 2500 Hz, `$\bigcirc$') involves the four modes until 2500 Hz. It shows good agreement with the full model, including the frequency band-gaps for the flexural waves at 850 Hz and 2000 Hz. 
	\begin{figure}[!ht]
  \centering
  \includegraphics[width=0.48\linewidth]{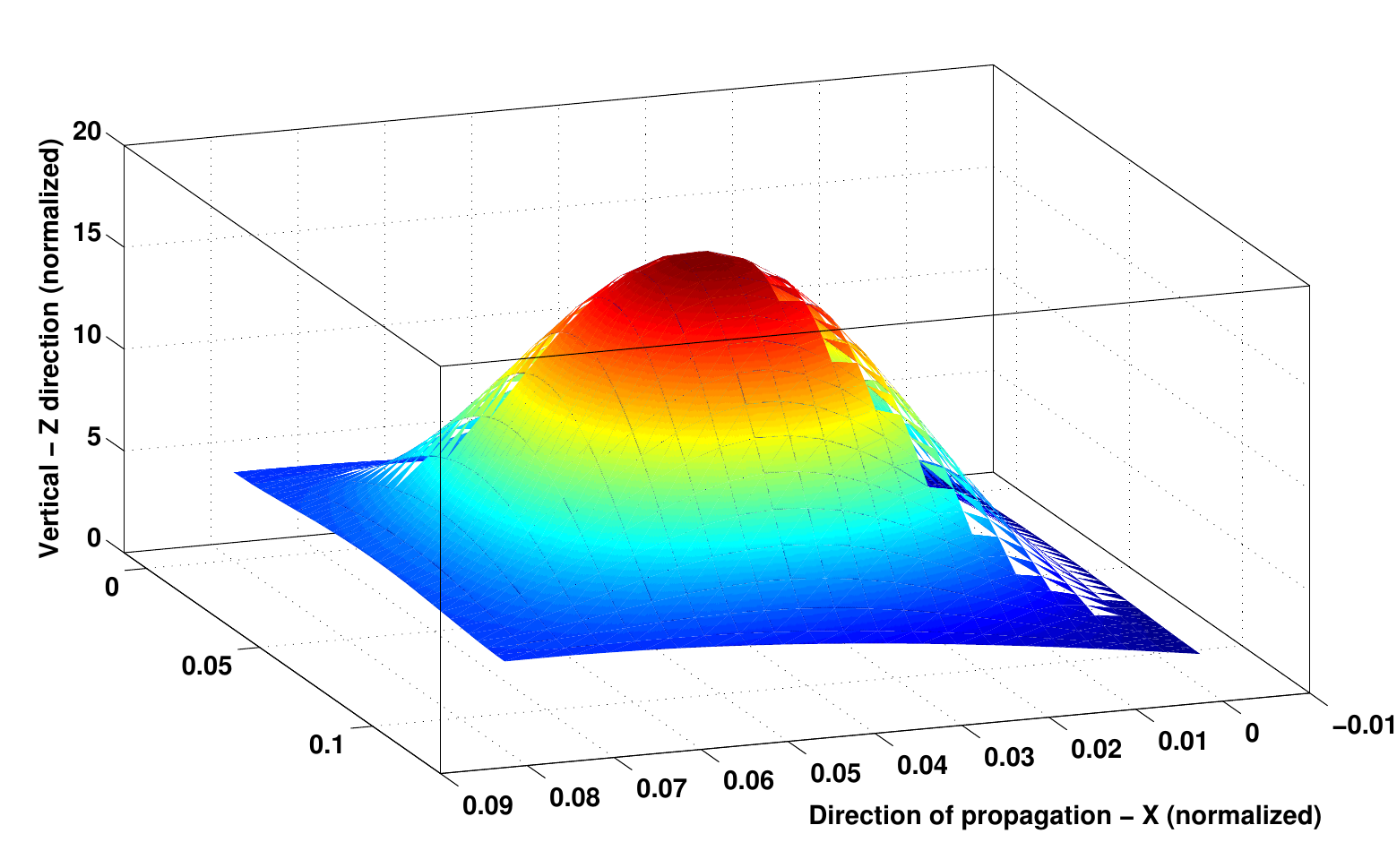}\hspace{0.03\linewidth}\includegraphics[width=0.48\linewidth]{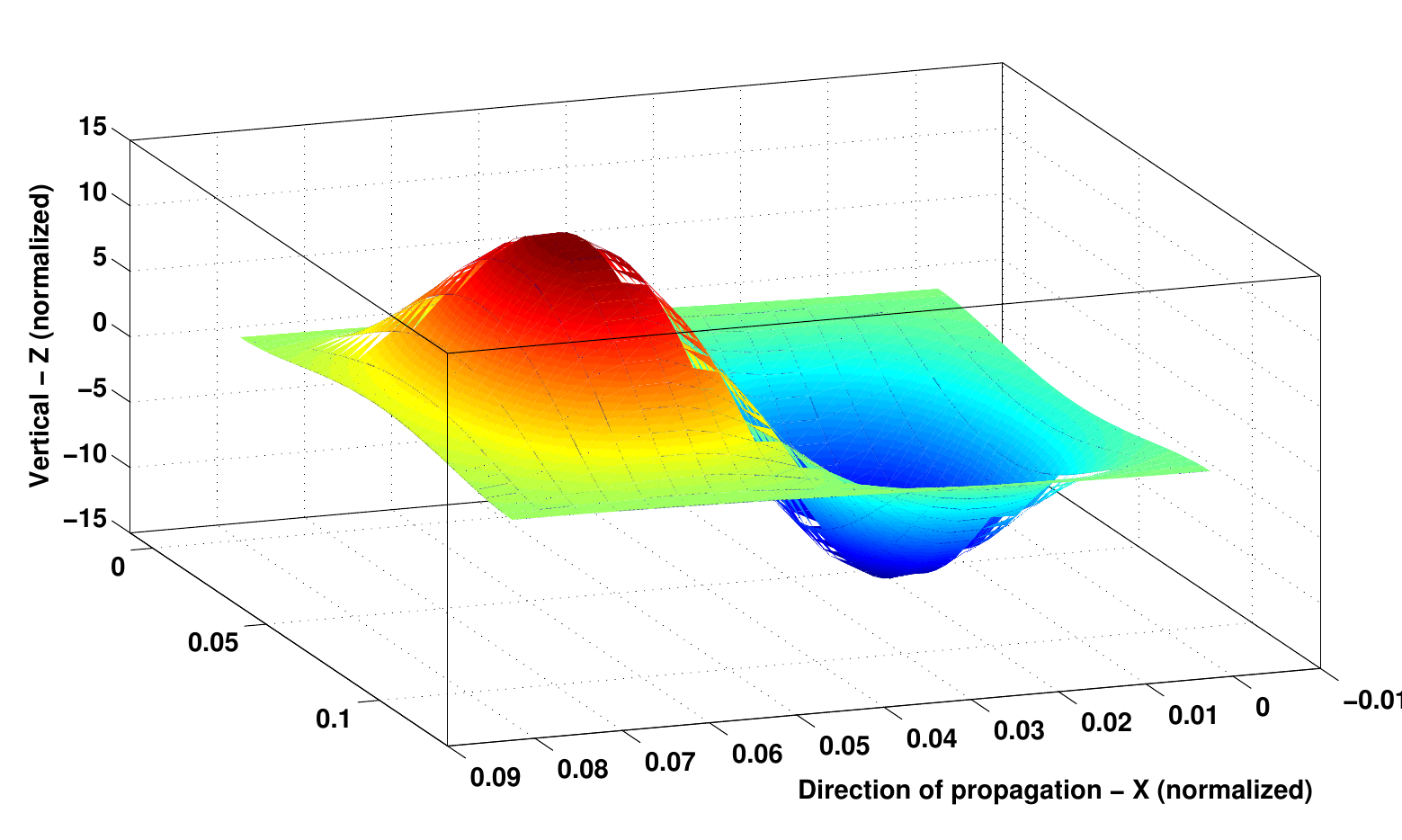}
  \caption{(a) Deformed shape of the periodic cell for the first local resonance at 800 Hz. (b) Deformed shape of the periodic cell for the second local resonance at 2000 Hz.\label{shape_mode}}
\end{figure}	
	The wave shapes associated with both local resonances at 800 Hz and 2000 Hz are shown in Fig.~\ref{shape_mode}. The second basis (CMS 1500 Hz, `$\times$') requires the two modes below 1500 Hz and accurately describes the first band-gap, but fails to predict the second band-gap at 2000 Hz. Similarly, the third basis (CMS 0 Hz, `$+$') is limited to the static constrained modes. Therefore the model is then limited to the low frequencies and fails to predict the band-gaps and high-order flexural waves. However it can be noticed that the longitudinal and shear waves are in good agreement in the three configurations. It can be explained since their shapes remain the same on the frequency range and can be described using the static constrained modes. The shape of the longitudinal wave at 1000 Hz and the shear wave at 750 Hz are shown respectively in Figs.~\ref{shape_long} and \ref{shape_shear}.

\subsubsection{Sensitivity of the method to wave-based reduction}
\begin{figure}[!ht]
  \centering
  \includegraphics[width=0.32\linewidth]{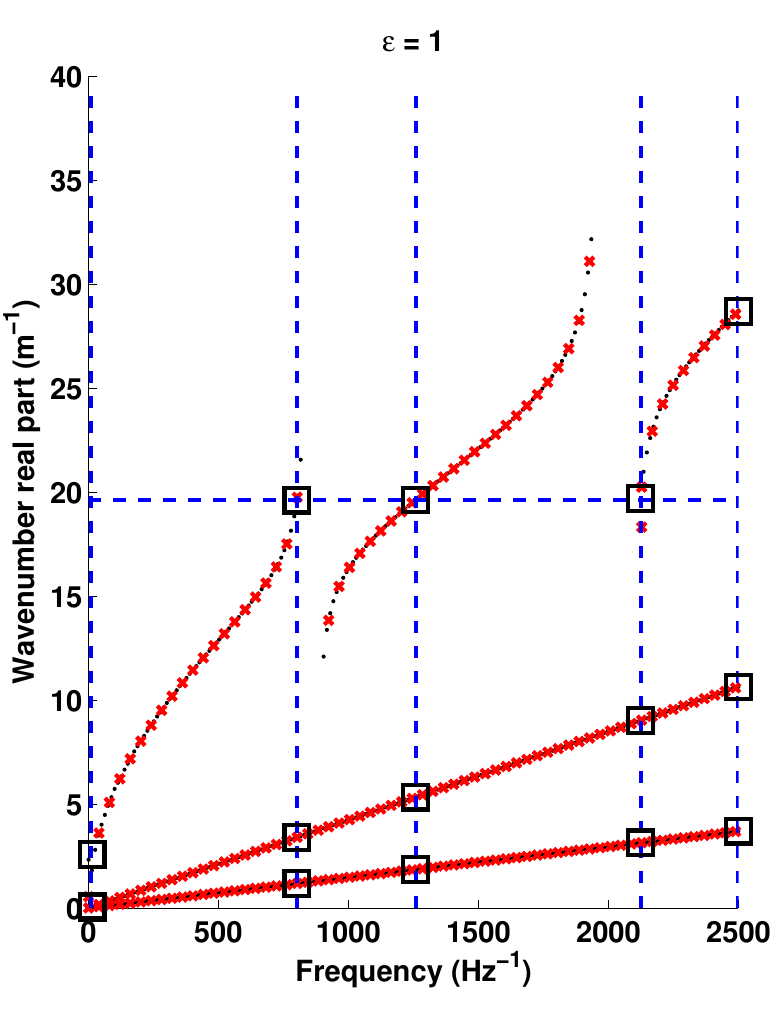}\hspace{0.01\linewidth}\includegraphics[width=0.32\linewidth]{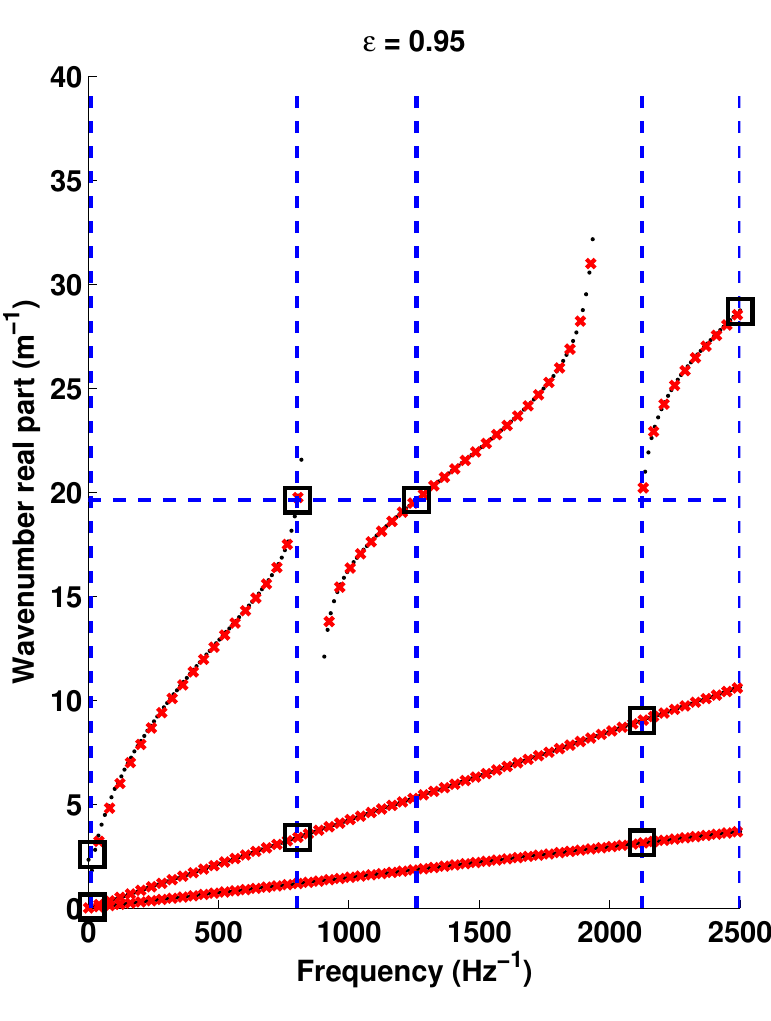}\hspace{0.01\linewidth}\includegraphics[width=0.32\linewidth]{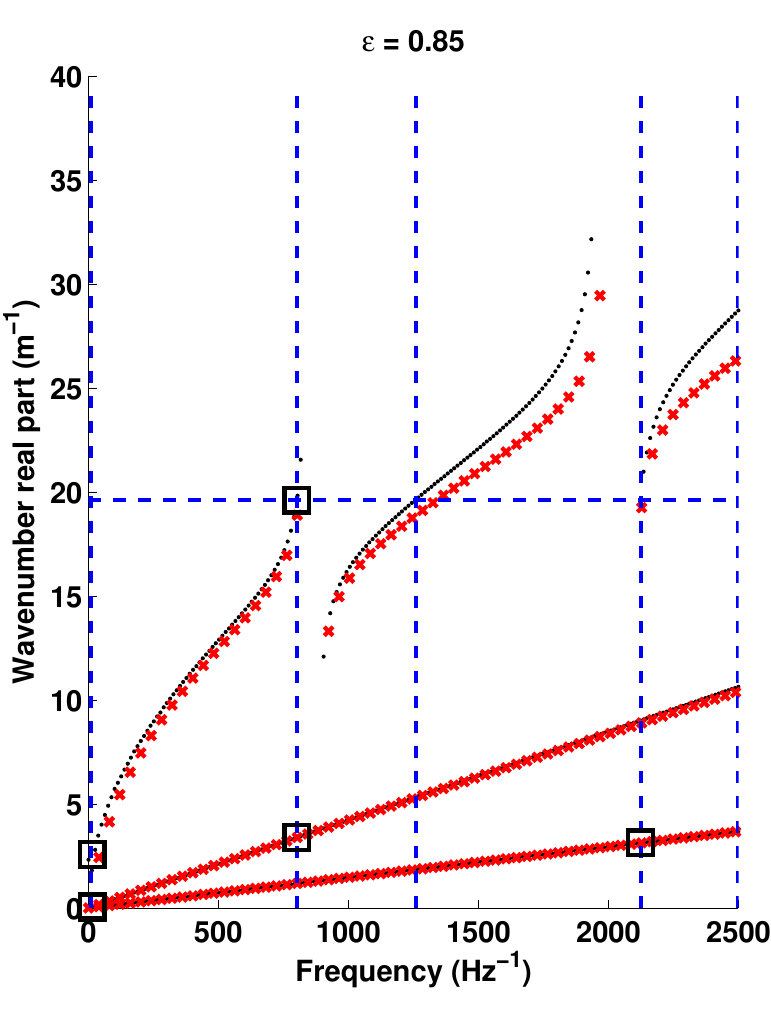}
  \caption{Influence of the wave selection criteria on the reduced model accuracy. The wavenumbers $k_y$ are determined for a given $k_x = 0$.}
  \label{epsilonfig}
\end{figure}

	The aforementioned wave-based interpolation involves a parameter $\varepsilon$ introduced in Eq.~(\ref{orthog}), to build the reduced set of shape vectors using a chosen subspace of the spectral problem solutions. The effect of this wave basis refinement is shown in Fig.~\ref{epsilonfig}. 
	In the proposed example, a choice of $\varepsilon = 0.85$ would considerably reduce the number of vectors retained in the shape basis but leads to an important error for the second and third orders of the flexural wave.
\begin{figure}[!ht]
  \centering
  \includegraphics[width=0.45\linewidth]{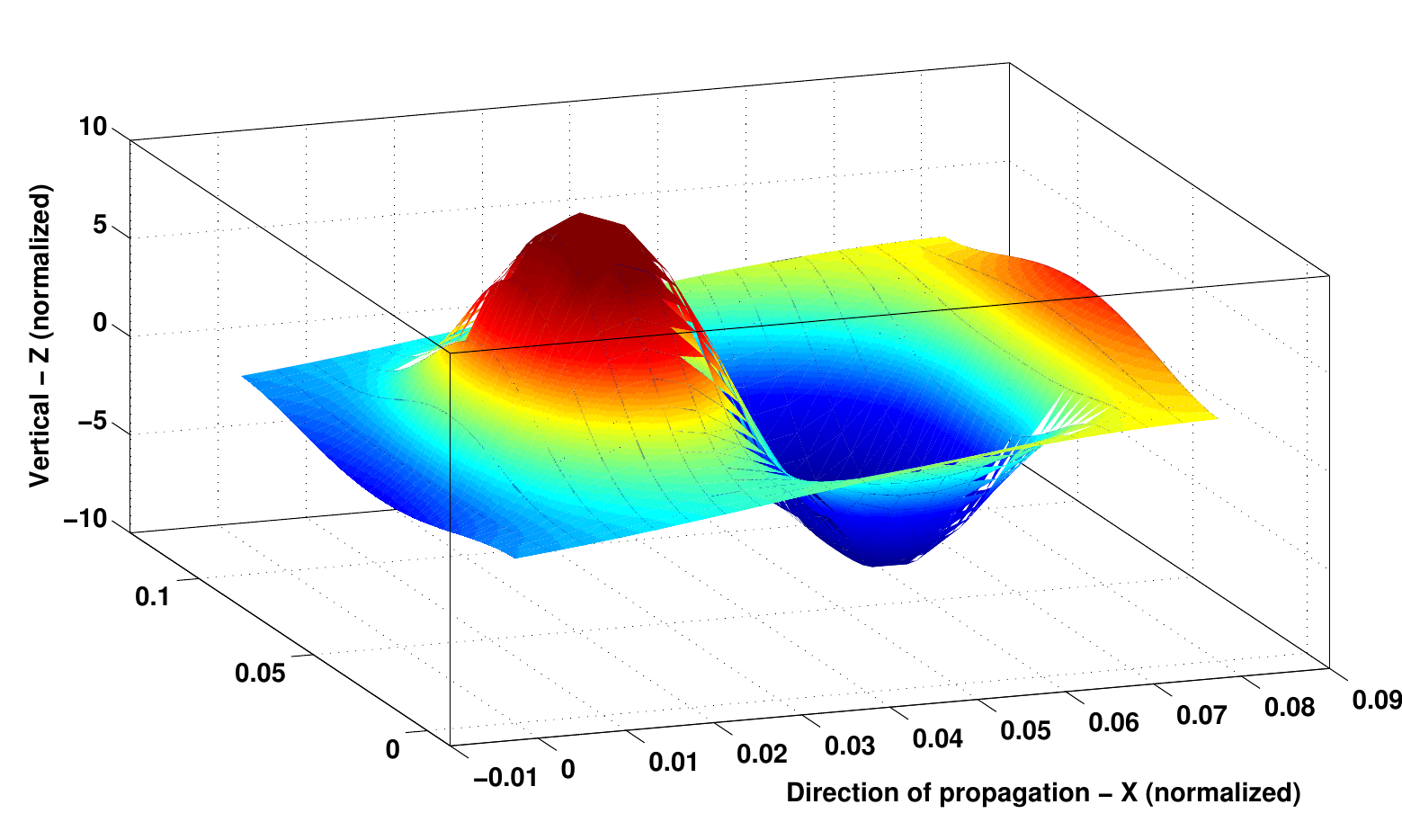}
  \caption{Deformed shape of the flexural wave at 2500 Hz.}
  \label{shape_2500hz}
\end{figure}
The third-order flexural wave shape is shown at 2500 Hz in Fig.~\ref{shape_2500hz}. Noteworthy, this displacement field cannot be described using the first and second orders flexural wave shapes. Oppositely, $\varepsilon = 1$ would mean that all of the propagating wave solutions are retained at each interpolation frequency. Therefore, it can be seen that the wave correlation parameter strongly influences the wave selection and the accuracy of the reduced model.

\section{Concluding remarks}\label{sec5}

The effectiveness of wave dispersion analysis for predicting the vibro-acoustic behaviour of large-scaled 2D waveguides led to numerous research work in the last decade. The Wave Finite Element Method (WFEM) is suitable tool for such purposes, but suffers numerical drawbacks throughout structurally advanced structures, industrial applications or for analysing localized wave propagation in broadband frequency ranges. In this paper, a formulation of the WFEM is described for 2D homogeneous or periodic waveguides, requiring large-scaled, structurally advanced or finely meshed finite element descriptions. This formulation is based on a two-scale model order reduction combining modal synthesis in the periodic cell with an innovative wave-based interpolation strategy, involving a projection of the spectral problem on a solution subspace associated with propagating wave shapes. The proposed formulation was applied for the k-space analysis of a bi-directionally stiffened panel on the frequency range $\left[ 0 - 2500 Hz \right]$. It provided very good results and a significant reduction of the computational cost. The stability of the method was investigated considering two key parameters of the wave-mode selection procedure: First, the influence of the fixed-interface modes used in the CMS to predict local resonances effects (important for stiffened panels) was highlighted. Noteworthy, these modes are essential to describe the inner cells displacement field, thereby the vibro-acoustic characteristics of the panel. Therefore, the authors recommend to select the modes between $0$ and $2 \times \omega_{max}$, where $\omega_{max}$ is the upper limit of the frequency range considered. Second, the effects of wave correlation criteria $\varepsilon$ were investigated to study the method's robustness. The proposed wave-based interpolation was successfully applied to reduce the spectral problem arising from Bloch's theorem, using propagating wave shapes participations on the cell's edges instead of the nodal displacements. A wave correlation criterion $\varepsilon$ of $0.99$ was chosen and provided accurate dispersion curves and significant reduction of the computational time. Besides, since most of the computational cost is associated to the propagating waves determination, an iterative procedure can be employed to enrich the wave basis to reach a required level of accuracy. It is hoped that it will provide an efficient numerical tool, encompassing numerical difficulties associated with the spectral analysis of structurally advanced 2D, continuous or periodic waveguides. An efficient prediction of local resonance effects and high-order propagating waves will provide useful results for the acoustic analysis of periodic panels.

\section*{Acknowledgements}
The authors would like to gratefully acknowledge Airbus Helicopters for their financial support.

\appendix
\numberwithin{equation}{section}
\renewcommand{\thesection}{Appendix~\Alph{section}}
\renewcommand{\theequation}{\Alph{section}.\arabic{equation}}

\newpage
\section{Determination of a reduced frequency subset using a fixed $(k_x,k_y)$ formulation}\label{annexe1}
	
	The following convention is used:
		\begin{equation}
	\mathbf{q}  = \left[\mathbf{q}_1^T \quad \mathbf{q}_2^T \quad \mathbf{q}_3^T \quad \mathbf{q}_4^T \quad \mathbf{q}_L^T \quad \mathbf{q}_B^T \quad \mathbf{q}_R^T \quad \mathbf{q}_T^T \quad \mathbf{P}_C^T \right]^T 
	= \left( \begin{array}{c}
	\mathbf{q}_b \\
	\mathbf{P}_C
	\end{array} \right)
	\end{equation}
	
	Then, using the periodicity relations defined Section~\ref{2.1}, the displacement vector involving modal participations can be written:
	\begin{equation}\label{lambdaR}
	\left(\begin{array}{c} 
	\mathbf{q}_b \\ 
	\mathbf{P}_C
	\end{array} \right)
	= \mathbf{\Lambda}_R' 
	\left( \begin{array}{c}
	\mathbf{q}_1 \\ 
	\mathbf{q}_L\\ 
	\mathbf{q}_B \\ 
	\mathbf{P}_C
	\end{array} \right)
	\end{equation} 
	
	with 
	\begin{equation}\label{lambdaR'}
	\mathbf{\Lambda}_R' = 
	\left( \begin{array}{cc}
	\mathbf{\Lambda}_R  & \mathbf{0} \\
	\mathbf{0} & \mathbf{I}_{\tt C} 
	\end{array}\right)
	\end{equation}
	
	where $\bf I_{\tt C}$ is an identity matrix of the size of $\mathbf{P}_C$, and $\mathbf{\Lambda}_R$ contains the propagation constants from the periodicity conditions:
	\begin{equation}
	\mathbf{\Lambda}_R = 
	\left[ \begin{array}{ccc} 
	\mathbf{I}_{\tt s} & \mathbf{0} & \mathbf{0} \\
	\lambda_x \mathbf{I}_{\tt s} & \mathbf{0} & \mathbf{0} \\
	\lambda_y \mathbf{I}_{\tt s} & \mathbf{0} & \mathbf{0} \\
	\lambda_x\lambda_y \mathbf{I} & \mathbf{0} & \mathbf{0} \\
	\mathbf{0} & \mathbf{I} & \mathbf{0} \\
	\mathbf{0} & \mathbf{0} & \mathbf{I} \\
	\mathbf{0} & \lambda_x \mathbf{I} & \mathbf{0} \\
	\mathbf{0} & \mathbf{0} & \lambda_y \mathbf{I}
	\end{array} \right]
	\end{equation}
	
	Note that the vector $\mathbf{P}_C$ is kept in the spectral eigenproblem since the frequency is unknown and no condensation of the inner DOFs can be done.	Similarly, the matrix $\mathbf{\Lambda}_L$ being defined as:
	\begin{equation}
	\mathbf{\Lambda}_L  \mathbf{f}_b = \mathbf{0}
	\end{equation}
	it yields:
	\begin{equation}\label{lambdaL'}
	\mathbf{\Lambda}_L'  =
	\left( \begin{array}{cc}
	\mathbf{\Lambda}_L  & \mathbf{0} \\
	\mathbf{0} & \mathbf{I}_{\tt C} 
	\end{array}\right)
	\end{equation}
	where 
	\begin{equation}
	\mathbf{\Lambda}_L = \left[ \begin{array}{cccccccc}
	\mathbf{I} & \lambda_x^{-1} \mathbf{I} & \lambda_y^{-1} \mathbf{I} & \lambda_x^{-1}\lambda_y^{-1} \mathbf{I} & \mathbf{0} & \mathbf{0} & \mathbf{0} & \mathbf{0} \\
	\mathbf{0} & \mathbf{0} & \mathbf{0} & \mathbf{0} & \mathbf{I} & \mathbf{0} & \lambda_x^{-1} \mathbf{I} & \mathbf{0} \\
	\mathbf{0} & \mathbf{0} & \mathbf{0} & \mathbf{0} & \mathbf{0} & \mathbf{I} & \mathbf{0} & \lambda_y^{-1} \mathbf{I}
	\end{array} \right]
	\end{equation}
	
	Therefore, the eigenproblem providing mid-aliasing frequency values: $k_x = \frac{\pi}{2 d_x}$, retained for the reduced solution subspace computation, can be written, with $\lambda_y$ given and $\lambda_x = exp(-j\dfrac{\pi}{2}) = -j$ as:
	\begin{equation}
	\left[ \mathbf{\Lambda}_L' (\lambda_x, \lambda_y) \mathbf{B}_\varPhi^T \left( \mathbf{\tilde{K}} - \omega_j^2 \mathbf{\tilde{M}}\right) \mathbf{B}_\varPhi \mathbf{\Lambda}_R'(\lambda_x, \lambda_y) \right] \mathbf{\Phi} = \mathbf{0}
	\end{equation}
	
\section{Formulation of the spectral eigenproblem for the 2D periodic waveguide}\label{annexe2}

	The condensed dynamic stiffness defined in Eq.~(\ref{condens}) can be ordered as follows:
	\begin{equation}\label{Dcond}
	\left[ 	\begin{array}{cccccccc} 	
	\mathbf{D}_{\tt 11}&\mathbf{D}_{\tt 12}&\mathbf{D}_{\tt 13}&\mathbf{D}_{\tt 14}&\mathbf{D}_{\tt 1L}&\mathbf{D}_{\tt 1B}&\mathbf{D}_{\tt 1R}&\mathbf{D}_{\tt 1T}\\
	\mathbf{D}_{\tt 21}&\mathbf{D}_{\tt 22}&\mathbf{D}_{\tt 23}&\mathbf{D}_{\tt 24}&\mathbf{D}_{\tt 2L}&\mathbf{D}_{\tt 2B}&\mathbf{D}_{\tt 2R}&\mathbf{D}_{\tt 2T}\\
	\mathbf{D}_{\tt 31}&\mathbf{D}_{\tt 32}&\mathbf{D}_{\tt 33}&\mathbf{D}_{\tt 34}&\mathbf{D}_{\tt 3L}&\mathbf{D}_{\tt 3B}&\mathbf{D}_{\tt 3R}&\mathbf{D}_{\tt 3T}\\
	\mathbf{D}_{\tt 41}&\mathbf{D}_{\tt 42}&\mathbf{D}_{\tt 43}&\mathbf{D}_{\tt 44}&\mathbf{D}_{\tt 4L}&\mathbf{D}_{\tt 4B}&\mathbf{D}_{\tt 4R}&\mathbf{D}_{\tt 4T}\\
	\mathbf{D}_{\tt L1}&\mathbf{D}_{\tt L2}&\mathbf{D}_{\tt L3}&\mathbf{D}_{\tt L4}&\mathbf{D}_{\tt LL}&\mathbf{D}_{\tt LB}&\mathbf{D}_{\tt LR}&\mathbf{D}_{\tt LT}\\
	\mathbf{D}_{\tt B1}&\mathbf{D}_{\tt B2}&\mathbf{D}_{\tt B3}&\mathbf{D}_{\tt B4}&\mathbf{D}_{\tt BL}&\mathbf{D}_{\tt BB}&\mathbf{D}_{\tt BR}&\mathbf{D}_{\tt BT}\\
	\mathbf{D}_{\tt R1}&\mathbf{D}_{\tt R2}&\mathbf{D}_{\tt R3}&\mathbf{D}_{\tt R4}&\mathbf{D}_{\tt RL}&\mathbf{D}_{\tt RB}&\mathbf{D}_{\tt RR}&\mathbf{D}_{\tt RT}\\
	\mathbf{D}_{\tt T1}&\mathbf{D}_{\tt T2}&\mathbf{D}_{\tt T3}&\mathbf{D}_{\tt T4}&\mathbf{D}_{\tt TL}&\mathbf{D}_{\tt TB}&\mathbf{D}_{\tt TR}&\mathbf{D}_{\tt TT}
	\end{array} \right] \left( \begin{array}{c}
	\mathbf{q}_1\\ \mathbf{q}_2\\ \mathbf{q}_3\\ \mathbf{q}_4\\ 
	\mathbf{q}_L\\ \mathbf{q}_B\\ \mathbf{q}_R\\ \mathbf{q}_T
	\end{array} \right) = \left( \begin{array}{c}
	\mathbf{f}_1\\ \mathbf{f}_2\\ \mathbf{f}_3\\ \mathbf{f}_4\\ 
	\mathbf{f}_L\\ \mathbf{f}_B\\ \mathbf{f}_R\\ \mathbf{f}_T 
	\end{array} \right)
	\end{equation}
	
	Using the matrices $\mathbf{\Lambda}_L$ and $\mathbf{\Lambda}_R$ defined in \ref{annexe1}, the product $\mathbf{\Lambda}_L (\lambda_x, \lambda_y) \, \mathbf{D}(\omega) \, \mathbf{\Lambda}_R(\lambda_x, \lambda_y)$ can be developed using the decomposition proposed for $\mathbf{D}$. Therefore, the coefficients $\mathbf{A}$, $\mathbf{B}$ and $\mathbf{C}$ are constant in a given $(\omega,\lambda_y)$ formulation.	Eq.~(\ref{spectralcond}) becomes a quadratic eigenvalue problem in $\lambda_x$, written:
	\begin{equation}
	\left(	\lambda _x\left[ \begin{array}{ccc}
	\mathbf{A}_{\tt 11} & \mathbf{A}_{\tt 1L} & \mathbf{A}_{\tt 1B} \\
	\mathbf{A}_{\tt L1} & \mathbf{A}_{\tt LL} & \mathbf{A}_{\tt LB} \\
	\mathbf{A}_{\tt B1} & \mathbf{A}_{\tt BL} & \mathbf{A}_{\tt BB}
	\end{array} \right] +
	\left[ \begin{array}{ccc}
	\mathbf{B}_{\tt 11} & \mathbf{B}_{\tt 1L} & \mathbf{B}_{\tt 1B} \\
	\mathbf{B}_{\tt L1} & \mathbf{B}_{\tt LL} & \mathbf{B}_{\tt LB} \\
	\mathbf{B}_{\tt B1} & \mathbf{B}_{\tt BL} & \mathbf{B}_{\tt BB}
	\end{array} \right] +
	\frac{1}{\lambda _x} \left[ \begin{array}{ccc}
	\mathbf{C}_{\tt 11} & \mathbf{C}_{\tt 1L} & \mathbf{C}_{\tt 1B} \\
	\mathbf{C}_{\tt L1} & \mathbf{C}_{\tt LL} & \mathbf{C}_{\tt LB} \\
	\mathbf{C}_{\tt B1} & \mathbf{C}_{\tt BL} & \mathbf{C}_{\tt BB}
	\end{array} \right] \right) 
	\left( \begin{array}{c}
	\mathbf{q}_1 \\ 
	\mathbf{q}_L \\ 
	\mathbf{q}_B
	\end{array} \right) = \mathbf{0}
	\end{equation}
	
	where $\lambda_x$ coefficients are:
	\begin{subequations}
	\begin{align}
	\mathbf{A}_{\tt 11}&=\mathbf{D}_{\tt 12} + \mathbf{D}_{\tt 34} + \mathbf{D}_{\tt 32}\lambda_y ^{-1} + \mathbf{D}_{\tt 14} \lambda_y \\
	\mathbf{A}_{\tt 1L}&=\mathbf{D}_{\tt 1R} + \mathbf{D}_{\tt 3R}\lambda_y^{-1} \\
	\mathbf{A}_{\tt L1}&=\mathbf{D}_{\tt L2} + \mathbf{D}_{\tt L4} \lambda_y \\
	\mathbf{A}_{\tt LL}&=\mathbf{D}_{\tt LR} \\
	\mathbf{A}_{\tt B1}&=\mathbf{D}_{\tt B2} + \mathbf{D}_{\tt T4} + \mathbf{D}_{\tt T2}\lambda_y^{-1} + \mathbf{D}_{\tt B4}\lambda_y \\
	\mathbf{A}_{\tt BL}&=\mathbf{D}_{\tt BR} + \mathbf{D}_{\tt TR} \lambda_y^{-1} \\
	\mathbf{A}_{\tt 1B}&=\mathbf{A}_{\tt LB} = \mathbf{A}_{\tt BB} = \mathbf{0}
	\end{align}
	\end{subequations}
	the constant coefficients are:
	\begin{subequations}
	\begin{align}
	\mathbf{B}_{\tt 11}&=\mathbf{D}_{\tt 11} + \mathbf{D}_{\tt 22} + \mathbf{D}_{\tt 33} + \mathbf{D}_{\tt 44} + (\mathbf{D}_{\tt 31} + \mathbf{D}_{\tt 42})\lambda_y^{-1} + (\mathbf{D}_{\tt 13}+\mathbf{D}_{\tt 24})\lambda_y \\
	\mathbf{B}_{\tt 1L}&=\mathbf{D}_{\tt 1L} + \mathbf{D}_{\tt 2R} + (\mathbf{D}_{\tt 3L} + \mathbf{D}_{\tt 4R})\lambda_y^{-1} \\
	\mathbf{B}_{\tt 1B}&=\mathbf{D}_{\tt 1B} + \mathbf{D}_{\tt 3T} + \mathbf{D}_{\tt 3B}\lambda_y^{-1} + \mathbf{D}_{\tt 1T}\lambda_y \\
	\mathbf{B}_{\tt L1}&=\mathbf{D}_{\tt L1} + \mathbf{D}_{\tt R2} + \mathbf{D}_{\tt L3}\lambda_y + \mathbf{D}_{\tt R4}\lambda_y \\
	\mathbf{B}_{\tt LL}&=\mathbf{D}_{\tt LL} + \mathbf{D}_{\tt RR} \\
	\mathbf{B}_{\tt LB}&=\mathbf{D}_{\tt LB} + \mathbf{D}_{\tt LT} \lambda_y \\
	\mathbf{B}_{\tt B1}&=\mathbf{D}_{\tt B1} + \mathbf{D}_{\tt T3} + \mathbf{D}_{\tt T1}\lambda_y^{-1} + \mathbf{D}_{\tt B3}\lambda_y \\
	\mathbf{B}_{\tt BL}&=\mathbf{D}_{\tt BL} + \mathbf{D}_{\tt TL} \lambda_y^{-1} \\
	\mathbf{B}_{\tt BB}&=\mathbf{D}_{\tt BB} + \mathbf{D}_{\tt TT} + \mathbf{D}_{\tt TB}\lambda_y^{-1} + \mathbf{D}_{\tt BT}\lambda_y
	\end{align}
	\end{subequations}
	
	and the coefficients in $\frac{1}{\lambda_x}$ are:
	\begin{subequations}
	\begin{align}
	\mathbf{C}_{\tt 11}&=\mathbf{D}_{\tt 21} + \mathbf{D}_{\tt 43} + \mathbf{D}_{\tt 41}\lambda_y^{-1} + \mathbf{D}_{\tt 23}\lambda_y \\
	\mathbf{C}_{\tt 1L}&=\mathbf{D}_{\tt 2L} + \mathbf{D}_{\tt 4L}\lambda_y^{-1} \\
	\mathbf{C}_{\tt 1B}&=\mathbf{D}_{\tt 2B} + \mathbf{D}_{\tt 4T} + \mathbf{D}_{\tt 4B}\lambda_y^{-1} + \mathbf{D}_{\tt 2T}\lambda_y \\
	\mathbf{C}_{\tt L1}&=\mathbf{D}_{\tt R1} + \mathbf{D}_{\tt R3}\lambda_y \\
	\mathbf{C}_{\tt LL}&=\mathbf{D}_{\tt RL} \\
	\mathbf{C}_{\tt LB}&=\mathbf{D}_{\tt RB} + \mathbf{D}_{\tt RT}\lambda_y \\
	\mathbf{C}_{\tt B1}&=\mathbf{C}_{\tt BL}=\mathbf{C}_{\tt BB}=\mathbf{0}
	\end{align}
	\end{subequations}


\begin{thebibliography}{32}
\providecommand{\natexlab}[1]{#1}
\providecommand{\url}[1]{\texttt{#1}}
\providecommand{\urlprefix}{URL }
\expandafter\ifx\csname urlstyle\endcsname\relax
  \providecommand{\doi}[1]{doi:\discretionary{}{}{}#1}\else
  \providecommand{\doi}{doi:\discretionary{}{}{}\begingroup
  \urlstyle{rm}\Url}\fi
\providecommand{\eprint}[2][]{\url{#2}}
\providecommand{\BIBand}{and}
\providecommand{\bibinfo}[2]{#2}
\ifx\xfnm\undefined \def\xfnm[#1]{\unskip,\space#1}\fi

\bibitem[{Mead et~al.(1988)Mead, Zhu and Bardell}]{MeadJSV1988}
D.J. Mead, D.C. Zhu, N.S. Bardell, Free vibration of an orthogonally stiffened flat plate, Journal of Sound and Vibration 127 (1988) 19--48.

\bibitem[{Mead(1973)}]{MeadJSV1973}
D.J. Mead, A general theory of harmonic wave propagation in linear periodic systems with multiple coupling, Journal of Sound and Vibration 27 (2) (1973) 235--260.

\bibitem[{Bloch(1929)}]{Bloch1929}
F. Bloch, Über die Quantenmechanik der Elektronen in Kristallgittern, Zeitschrift für Physik 52 (7--8) (1929) 555--600.

\bibitem[{Mejdi and Atalla(2010)}]{MejdiAtallaJSV2010}
A. Mejdi, N. Atalla, Dynamic and acoustic response of bidirectionally stiffened plates with eccentric stiffeners subject to airborne and structure-borne excitations, Journal of Sound and Vibration 329 (21) (2010) 4422--4439.

\bibitem[{Mejdi and Atalla(2012)}]{MejdiAtallaIJMS2012}
A. Mejdi, N. Atalla, Vibroacoustic analysis of laminated composite panels stiffened by complex laminated composite stiffeners, International Journal of Mechanical Sciences 58 (2012) 13--26.

\bibitem[{Sutyrin(1997)}]{SutyrinAM1997}
V. Sutyrin, Derivation of plate theory accounting asymptotically correct shear deformation, Journal of Applied Mechanics 64 (4) (1997) 905--915.

\bibitem[{Chronopoulos et~al.(2013)Chronopoulos, Troclet, Bareille and
  Ichchou}]{ChronopoulosCS2013}
D. Chronopoulos, B. Troclet, O. Bareille, M. Ichchou, Modeling the response of composite panels by a dynamic stiffness approach, Composite Structures 96 (2013) 111--120.

\bibitem[{Ruzzene et~al.(2003)Ruzzene, Scarpa and Soranna}]{RuzzeneSMS2003}
M. Ruzzene, F. Scarpa, F. Soranna, Wave beaming effects in two-dimensional cellular structures, Smart Materials and Structures 12 (3) (2003) 363.

\bibitem[{Lyon et~al.(1995)Lyon, DeJong and Heckl}]{Lyon1995}
R.H. Lyon, R.G. DeJong, M. Heckl, Theory and application of statistical energy analysis, Journal of the Acoustical Society of America 98 (1995) 3021.

\bibitem[{Mace et~al.(2005)Mace, Duhamel, Brennan and Hinke}]{MaceJASA2005}
B.R. Mace, D. Duhamel, M.J. Brennan, L. Hinke, Finite element prediction of wave motion in structural waveguides, Journal of the Acoustical Society of America 117 (2005) 2835--2843.

\bibitem[{Mencik and Ichchou(2005)}]{MencikEJM2005}
J.M. Mencik, M.N. Ichchou, Multi-mode propagation and diffusion in structures through finite elements, European Journal of Mechanics A---Solid 24 (5) (2005) 877--898.

\bibitem[{W.~X.~Zhong(1995)}]{Zhong1995}
W.X. Zhong, F.W. Williams, On the direct solution of wave propagation for repetitive structures, Journal of Sound and Vibration 181 (1995) 485--501.

\bibitem[{Ichchou et~al.(2008)Ichchou, Berthaut and Collet}]{IchchouIJSS2008}
M.N. Ichchou, J. Berthaut, M. Collet, Multi-mode wave propagation in ribbed plates: Part I, wavenumber-space characteristics; Part II, predictions and comparisons, International Journal of Solids and Structures 45 (5) (2008) 1179--1216.

\bibitem[{Houillon et~al.(2005)Houillon, Ichchou and
  Jezequel}]{HouillonJSV2005}
L. Houillon, M.N. Ichchou, L. Jezequel, Wave motion in thin walled structures, Journal of Sound and Vibration 281 (2005) 483--507.

\bibitem[{Waki et~al.(2009{\natexlab{a}})Waki, Mace and Brennan}]{WakiJSV2009a}
Y. Waki, B. Mace, M. Brennan, Free and forced vibrations of a tyre using a wave/finite element approach, Journal of Sound and Vibration 323 (3--5) (2009) 737--756.

\bibitem[{Chronopoulos et~al.(2014)Chronopoulos, Ichchou, Troclet and
  Bareille}]{ChronopoulosCS2014}
D. Chronopoulos, M.N. Ichchou, B. Troclet, O. Bareille, Predicting the broadband response of a layered cone--cylinder--cone shell, Composite Structures 107 (2014) 149--159.

\bibitem[{Mace and Manconi(2008)}]{MaceManconiJSV2008}
B.R. Mace, E. Manconi, Modelling wave propagation in two-dimensional structures using finite element analysis, Journal of Sound and Vibration 318 (4) (2008) 884--902.

\bibitem[{Collet et~al.(2011)Collet, Ouisse, Ruzzene and
  Ichchou}]{ColletIJSS2011}
M. Collet, M. Ouisse, M. Ruzzene, M. Ichchou, Floquet--Bloch decomposition for the computation of dispersion of two-dimensional periodic, damped mechanical systems, International Journal of Solids and Structures 48 (20) (2011) 2837--2848.

\bibitem[{Orrenius et~al.(2013)Orrenius, Liu, Wareing, Finnveden and
  Cotoni}]{OrreniusWM2013}
U. Orrenius, H. Liu, A. Wareing, S. Finnveden, V. Cotoni, Wave modelling in predictive vibro-acoustics: applications to rail vehicles and aircraft, Wave Motion 51 (4) (2013) 635--649.

\bibitem[{Renno and Mace(2011)}]{Renno2011}
J. Renno, B. Mace, Calculating the forced response of two-dimensional homogeneous media using the wave and finite element method, Journal of Sound and Vibration 330 (2011) 5913--5927.

\bibitem[{Waki et~al.(2009{\natexlab{b}})Waki, Mace and Brennan}]{WakiJSV2009b}
Y. Waki, B. Mace, M. Brennan, Numerical issues concerning the wave and finite element method for free and forced vibrations of waveguides, Journal of Sound and Vibration 327 (1--2) (2009) 92--108.

\bibitem[{Mencik(2012)}]{MencikCMAME2012}
J.M. Mencik, A model reduction strategy for computing the forced response of elastic waveguides using the wave finite element method, Computer Methods in Applied Mechanics and Engineering 229--232 (2012) 68--86.

\bibitem[{J.~M.~Renno(2010)}]{Renno2010}
J.M. Renno, B.R. Mace, On the forced response of waveguides using the wave and finite element method, Journal of Sound and Vibration 329 (2010) 5474--5488.

\bibitem[{Droz et~al.(2014)Droz, Lain{\'e}, Ichchou and
  Inqui{\'e}t{\'e}}]{DrozCS2014}
C. Droz, J.P. Lainé, M. Ichchou, G. Inquiété, A reduced formulation for the free-wave propagation analysis in composite structures, Composite Structures 113 (2014) 134--144.

\bibitem[{Langley(1993)}]{LangleyJSV1993}
R. Langley, A note on the force boundary conditions for two-dimensional periodic structures with corner freedoms, Journal of Sound and Vibration 167 (2) (1993) 377--381.

\bibitem[{Akrout(2005)}]{AkroutPhD2005}
S. Akrout. Comportement dynamique déterministe et large bande des structures guidées, PhD thesis, Ecole Centrale de Lyon, 2005.

\bibitem[{Cotoni et~al.(2008)Cotoni, Langley and Shorter}]{CotoniJSV2008}
V. Cotoni, R. Langley, P. Shorter, A statistical energy analysis subsystem formulation using finite element and periodic structure theory, Journal of Sound and Vibration 318 (4) (2008) 1077--1108.

\bibitem[{Langley(1997)}]{Langley1997}
R. Langley, Some perspectives on wave-mode duality. In: IUTAM Symposium on Statistical Energy Analysis, 1997, pp. 1--12.

\bibitem[{Craig and Bampton(1968)}]{Craig1968}
R. Craig, M. Bampton, Coupling of substructures for dynamic analyses, AIAA 6 (7) (1968) 1313--1319.

\bibitem[{Mead and Parthan(1979)}]{MeadJSV1979}
D. Mead, S. Parthan, Free wave propagation in two-dimensional periodic plates, Journal of Sound and Vibration 64 (3) (1979) 325--348.

\bibitem[{Orrenius and Finnveden(1996)}]{Orrenius1996}
U. Orrenius, S. Finnveden, Calculation of wave propagation in rib-stiffened plate structures, Journal of Sound and Vibration 198 (1996) 203--224.

\bibitem[{Legault et~al.(2011)Legault, Mejdi and N}]{Legault2011}
J. Legault, A. Mejdi, N. Atalla., Vibro-acoustic response of orthogonally stiffened panels: the effects of finite dimensions, Journal of Sound and Vibration 183 (2011) 5928--5948.

\end{thebibliography}
\end{document}